\documentclass[journal]{IEEEtran}
\usepackage{graphicx}
\usepackage{tipa}
\usepackage{bbm}
\usepackage{mathrsfs}
\usepackage{amsfonts}
\usepackage{cite,url,subfigure,epsfig,graphicx}
\usepackage{amssymb,amsmath}
\usepackage{indentfirst}
\usepackage[linesnumbered,ruled,vlined]{algorithm2e}
\usepackage{pdfpages}
\usepackage{multirow}
\usepackage{cases}
\usepackage{cite,url,subfigure,epsfig,graphicx}
\usepackage{times,verbatim,amsfonts,amsmath,color}
\usepackage{float}
\IEEEoverridecommandlockouts
\makeatletter
\def\ps@headings{\def\@oddhead{\mbox{}\scriptsize\rightmark \hfil \thepage}\def\@evenhead{\scriptsize\thepage \hfil \leftmark\mbox{}}\def\@oddfoot{}\def\@evenfoot{}}
\makeatother 

\newtheorem{definition}{ \textbf{Definition}}
\newtheorem{lemma}{ \textbf{Lemma}}

\newtheorem{theorem}{ \textbf{Theorem}}

\newtheorem{problem}{ \textbf{Problem}}

\newtheorem{corollary}{ \textbf{Corollary}}

\begin{document}

\title{Collaborative Honeypot Defense in UAV Networks: A Learning-Based Game Approach}
\author{
\IEEEauthorblockN{Yuntao~Wang\IEEEauthorrefmark{2}, Zhou~Su\IEEEauthorrefmark{2}\IEEEauthorrefmark{1}, Abderrahim~Benslimane\IEEEauthorrefmark{3}, Qichao~Xu\IEEEauthorrefmark{5}, Minghui~Dai\IEEEauthorrefmark{4}, and Ruidong~Li\IEEEauthorrefmark{9}}\\
\IEEEauthorblockA{
\IEEEauthorrefmark{2}School of Cyber Science and Engineering, Xi'an Jiaotong University, China\\
\IEEEauthorrefmark{3}Laboratory of Computer Sciences, Avignon University, France\\
\IEEEauthorrefmark{5}School of Mechatronic Engineering and Automation, Shanghai University, China\\
\IEEEauthorrefmark{4}State Key Laboratory of Internet of Things for Smart City, University of Macau, Macau, China\\
\IEEEauthorrefmark{9}Department of Electrical and Computer Engineering, Kanazawa University, Japan\\
\IEEEauthorrefmark{1}Corresponding Author: zhousu@ieee.org
}
\thanks{Manuscript accepted Aug. 28, 2023 by IEEE Transactions on Information Forensics \& Security. This work has been partly presented at IEEE GLOBECOM2022 \cite{10000872}.}}

\maketitle

\begin{abstract}
The proliferation of unmanned aerial vehicles (UAVs) opens up new opportunities for on-demand service provision anywhere and anytime, but also exposes UAVs to a variety of cyber threats. Low/medium interaction honeypots offer a promising lightweight defense for actively protecting mobile Internet of things, particularly UAV networks.
While previous research has primarily focused on honeypot system design and attack pattern recognition, the incentive issue for motivating UAV's participation (e.g., sharing trapped attack data in honeypots) to collaboratively resist distributed and sophisticated attacks remains unexplored.
This paper proposes a novel game-theoretical collaborative defense approach to address optimal, fair, and feasible incentive design, in the presence of network dynamics and UAVs' multi-dimensional private information (e.g., valid defense data (VDD) volume, communication delay, and UAV cost).
Specifically, we first develop a honeypot game between UAVs and the network operator under both partial and complete information asymmetry scenarios. The optimal VDD-reward contract design problem with partial information asymmetry is then solved using a contract-theoretic approach that ensures budget feasibility, truthfulness, fairness, and computational efficiency.
In addition, under complete information asymmetry, we devise a distributed reinforcement learning algorithm to dynamically design optimal contracts for distinct types of UAVs in the time-varying UAV network.
Extensive simulations demonstrate that the proposed scheme can motivate UAV's cooperation in VDD sharing and improve defensive effectiveness, compared with conventional schemes.
\end{abstract}


\IEEEpeerreviewmaketitle
%----------------------------------------------------------------------------------
\section{Introduction}
With the advancements in communication and embedded technologies, unmanned aerial vehicles (UAVs) have been widely employed in a variety of applications including power lines inspection, medical delivery, disaster search, and crowd surveillance \cite{8943319,9696188,9123863}.
Thanks to their low cost, 3D mobility, and flexible deployment, UAVs can be swiftly dispatched to hard-to-reach sites (e.g., disaster zones) to undertake time-critical missions and offer urgent communication services using line-of-sight (LoS) links \cite{9453853,9035635,10221755}.
As UAVs are computer-controlled agents with wireless/radio interfaces, the widespread use of UAVs in service offering exposes them to a plethora of sophisticated cyberattacks\footnotemark[1] \cite{8946587} such as denial-of-service (DoS), hijacking, and data theft.

\begin{figure}[!t]\setlength{\abovecaptionskip}{-0.2cm}
\centering
  \includegraphics[height=3.2cm]{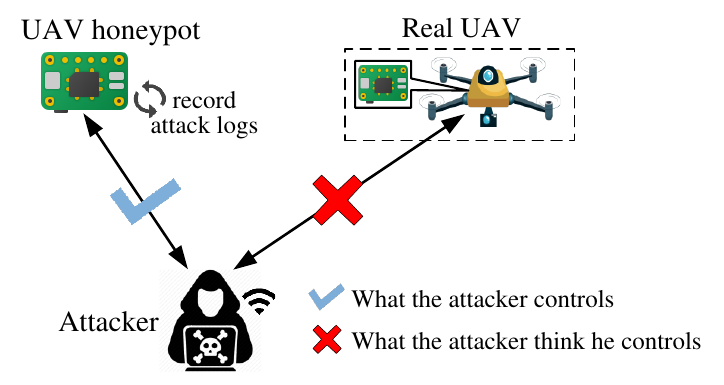}%width=6.2cm,
  \caption{{An illustrating example of active UAV defense via the portable UAV honeypot.}}\label{fig:honeypot-UAV}\vspace{-5.mm}
\end{figure}

In the face of escalating cyber threats, low/medium-interaction honeypots, as a supplemental active defense technology, provide a cost-effective alternative to strengthen UAV defense \cite{9359711,6587320,4267549}. Honeypots are physical or virtual systems that imitate real devices to lure and trap intruders, allowing defenders to continuously learn new attack patterns \cite{9500567,10090432}.
Low/medium-interaction honeypots (which simulate network operations on the TCP/IP stack) can provide lightweight defenses, compared to resource-hungry high-interaction honeypots \cite{4267549,9149314}. These defenses are particularly suitable for mobile and resource-constrained devices (such as battery-powered UAVs), which have drawn numerous research efforts.
For example, Vasilomanolakis \emph{et al}. \cite{Vasi2014HosTaGe} develop the HosTaGe prototype, a generic low-interaction honeypot, for mobile devices to identify fraudulent wireless networks as they connect.
Meanwhile, a medium-interaction honeypot prototype called HoneyDrone is implemented on small-size UAVs in \cite{Daub2018HoneyDrone} via simulating UAV-specific protocols in the designed UAV honeypot.
{Motivated by existing works, we leverage the portable and lightweight UAV honeypots to protect real UAVs, by emulating UAV's radio protocols and trapping cyberattackers in the honeypots, as shown in Fig.~\ref{fig:honeypot-UAV}. Besides, the logs of attackers' behaviors can be captured by the honeypots to help learn the attack patterns and design the UAV defense strategy.}\footnotetext[1]{{As flying UAVs can be detected and monitored by multiple sensors (e.g., advanced carema and radar), various physical-layer attacks (e.g., jamming, eavesdropping, and GPS spoofing) can also threaten and even capture the target UAV. The \emph{physical-layer} attacks on UAVs are not the scope of this work, and interested readers can refer to \cite{8883127,9585491}.}}

Despite the fundamental contributions to system and software design of existing literature \cite{Vasi2014HosTaGe,Daub2018HoneyDrone}, the honeypot-based cooperative defensive strategy for UAVs is rarely studied.
Particularly, given the current trend of distributed, sophisticated, and complex covert cyber attacks (e.g., advanced persistent threat (APT) and distributed DoS (DDoS)), there is a necessity for large-scale collaborative defense among UAVs for global situational awareness by exchanging trapped attack information (e.g., attack interaction logs) in local honeypots.
Nonetheless, as participating in such collaboration mechanisms entails significant costs (e.g., honeypot execution and communication costs) and potential privacy leakage (e.g., UAV configuration and flying route), UAVs might be reluctant to share their captured attack data without adequate incentives. Additionally, malicious UAVs may distribute false attack information to mislead others. Therefore, it is imperative to design an effective incentive mechanism to encourage UAVs to honestly cooperate in the joint defense.

However, the following key challenges need to be resolved to design such an incentive mechanism compatible with UAVs. First, UAVs typically have multi-dimensional private information in terms of valid defense data (VDD) volume, communication delay, VDD cost, and privacy cost.
Moreover, selfish UAVs may launch free-riding attacks, namely, they will not contribute to but still benefit from the joint defense, thereby disincentivizing honest UAVs. The presence of UAV's multi-dimensional information asymmetry and free-riding behaviors poses significant challenges in \emph{optimally} and \emph{fairly} distributing rewards to compensate for {UAVs'} costs.
Besides, as both UAV networks and attack behaviors can be highly dynamic, the shared defense data from UAVs should be timely aggregated to produce real-time defense strategies. As such, it remains a challenge to \emph{feasibly} implement the incentive mechanism in practical UAV applications with time-varying environments and stringent latency requirements.

\begin{figure}[!t]\setlength{\abovecaptionskip}{-0.15cm}
\centering
  \includegraphics[width=8.0cm]{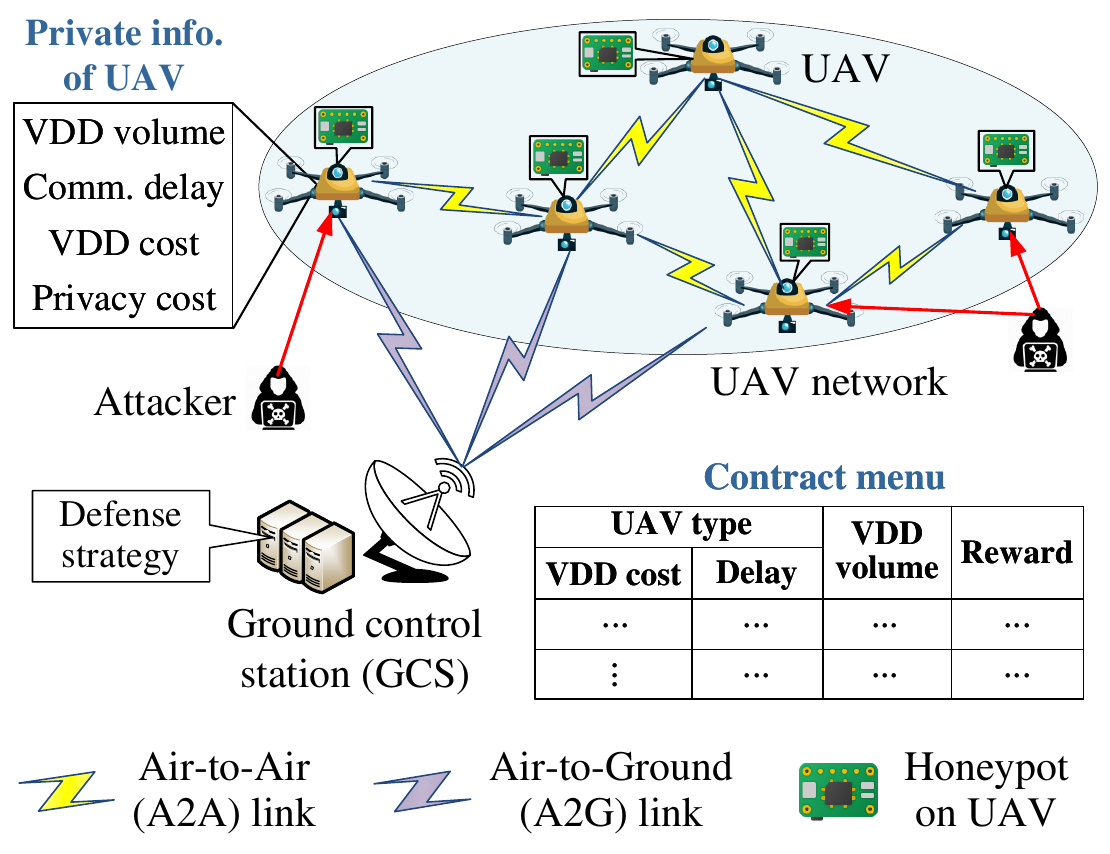}
  \caption{Illustration of the incentive-driven honeypot game for collaborative defense via sharing VDD in UAVs' local honeypots.}\label{fig:model}\vspace{-4.5mm}
\end{figure}

To address these issues, this paper proposes a novel incentive-driven honeypot-based collaborative UAV defense scheme to enhance defensive effectiveness, in which optimal, fair, and feasible incentives are offered to promote UAVs' honest cooperation in the face of information asymmetry and network dynamics.
Firstly, we present a UAV honeypot game{-based} framework consisting of multiple UAVs and a ground control station (GCS) serving as the network operator. In the game, as shown in Fig.~\ref{fig:model}, the GCS designs a series of contracts (specifying the relation among VDD size, VDD cost, communication delay, and rewards) for heterogeneous UAVs, and each UAV chooses a contract to share its defense data.
Then, we formulate the optimal contract design problem for the GCS under practical constraints and different levels of information asymmetry. Next, by leveraging the revelation principle, we analytically derive the optimal contract under \emph{partial information asymmetry} (i.e., the GCS only knows the distribution of UAVs' private types) and rigorously prove its truthfulness, fairness, budget feasibility (BF), and computational efficiency.
Furthermore, under \emph{complete information asymmetry} (i.e., {the GCS doesn't know any information of UAVs' private types even for its distribution}), we develop a two-tier reinforcement learning (RL) algorithm to dynamically acquire optimal contracts for different types of UAVs via trials with high adaption to the fast-changing network environment.

The main contribution of this work is four-fold as below.
\begin{itemize}
  \item \emph{Honeypot game theoretical cooperative defense framework}. We propose an active and cooperative defense framework based on honeypot game to encourage distributed UAVs to share honeypot data with the defensive designer (i.e., the GCS). Under this framework, {a} feasible incentive {mechanism is} designed to forbid free-riding UAVs while ensuring compensation fairness and {utility} optimality under different levels of information asymmetry.
  \item \emph{Budget-constrained optimal contract design under partial information asymmetry}. We leverage the multi-dimensional contract theory to design optimal fair contracts for heterogeneous UAVs with multi-dimensional private types under partial incomplete information. By summarizing UAV's multi-dimensional private type into a one-dimensional criterion, the optimal data-payment contract is theoretically solved. Besides, an adaptive dynamic assignment algorithm is designed for practical deployment under budget constraints.
  \item \emph{RL-based optimal contract design under complete information asymmetry}. By formulating UAVs' and GCS's interactions as finite Markov decision processes (MDPs), we devise the distributed policy hill-climbing (PHC) algorithm with two tiers to dynamically learn the optimal contractual strategies of all participants under strongly incomplete information. A hotbooting method is also designed in PHC learning to accelerate convergence rate by initializing the Q-values and mixed-strategy tables using historical experience.
  \item \emph{Extensive simulations for performance evaluation}. We evaluate the efficiency and effectiveness of the proposed scheme using extensive simulations. Numerical results show that the proposed scheme can effectively defend against free-riders and motivate UAVs' participation in honeypot defense with improved UAV utility and defensive effectiveness in both partial and complete information asymmetry scenarios, in comparison to conventional approaches.
\end{itemize}

\begin{table*}[!t]\setlength{\abovecaptionskip}{-0.0cm}
\begin{center}
\caption{Existing representative game-theoretical honeypot defense approaches: A comparative summary}\label{Table1}
\begin{tabular}{|c|c|c|cc|c|c|}
\hline
\multirow{2}{*}{Ref.} & \multirow{2}{*}{Scenario} & \multirow{2}{*}{\begin{tabular}[c]{@{}c@{}}Mobile \\ honeypot\end{tabular}} & \multicolumn{2}{c|}{User's private info.}    & \multirow{2}{*}{\begin{tabular}[c]{@{}c@{}}Budget\\ feasibility\end{tabular}}    & \multirow{2}{*}{\begin{tabular}[c]{@{}c@{}}Partial \& complete information \\ asymmetry scenarios\end{tabular}} \\ \cline{4-5}
                      &                     &                                                        & \multicolumn{1}{c|}{cost}                  & delay               &      &                                                                                                \\ \hline
{\cite{4267549}}      &$N.A.$        & $\times$                                                                    & \multicolumn{1}{c|}{$\checkmark$}          & $\times$              & $\times$    & $\times$                                                                                          \\ \hline
{\cite{9431233}}      &Industrial IoT      & $\times$                                                                    & \multicolumn{1}{c|}{$\checkmark$}          & $N.A.$              & $\times$      & $\times$                                                                                        \\ \hline
{\cite{7857804}}      &Power grid      & $\times$                                                                    & \multicolumn{1}{c|}{$\checkmark$}          & $N.A.$              & $\times$      & $\times$                                                                                        \\ \hline
{\cite{9397770}}      &Power grid        & $\times$                                                                    & \multicolumn{1}{c|}{$\checkmark$}          & $\times$              & $\times$    & $\times$                                                                                          \\ \hline
{\cite{7442780,9435800}}      &General IoT        & $\times$                                                                    & \multicolumn{1}{c|}{$\checkmark$}          & $\times$              & $\times$    & $\times$                                                                                          \\ \hline
{\cite{8675527}}      &Cloud computing      & $\times$                                                                    & \multicolumn{1}{c|}{$\checkmark$}          & $N.A.$              & $\times$      & $\times$                                                                                        \\ \hline
\textbf{Ours}         &UAV network & \textbf{$\checkmark$}                                                       & \multicolumn{1}{c|}{\textbf{$\checkmark$}} & \textbf{$\checkmark$} & \textbf{$\checkmark$}  & \textbf{$\checkmark$}                                                                              \\ \hline
\end{tabular}
\end{center}
\end{table*}

The remainder of the paper is organized as follows. Section \ref{sec:RELATEDWORK} reviews the related works. Section \ref{sec:SYSTEMMODEL} introduces the system model and Section \ref{sec:GAME} formulates the honeypot game-based cooperative defense framework. Section \ref{sec:CompleteSolution} designs the optimal contract under the ideal complete information {scenario}. Section \ref{sec:StaticSolution} and Section \ref{sec:DynamicSolution} present contract-based and RL-based optimal incentive mechanisms under partial and complete information asymmetry scenarios, respectively. Performance evaluation is given in Section \ref{sec:SIMULATION} and conclusions are drawn in Section \ref{sec:CONSLUSION}.

\section{Related Works}\label{sec:RELATEDWORK}
In this section, we review the related literature on honeypot-based defenses in the Internet of things (IoT) and UAV networks, as well as game modeling for honeypot-based defense.

\vspace{-2mm}\subsection{Honeypot-Based Defenses in IoT and UAV Networks}\label{subsec:RELATEDWORK1}
Honeypot offers an active line of defense for the IoT and mobile UAV networks by trapping and deceiving cyber attackers via carefully monitored unprotected systems.
The level of interaction with adversaries can be used to classify honeypots \cite{4267549}. High-interaction honeypots are real hosts or virtual machines (VMs) that replicate all of the functionalities of a real system, which are resource-hungry and typically expensive to maintain.
While low/medium-interaction honeypots only simulate the networking stack at a low/medium granularity to provide detailed logging and monitoring functionalities, which are much cheaper to maintain and better suited for mobile devices (e.g., UAVs) \cite{Vasi2014HosTaGe,Daub2018HoneyDrone}.
Fan \emph{et al}. \cite{8635491} present a novel all-round high-interaction honeypot system to efficiently acquire high-quality attack data in the large-scale IoT. By decoupling the honeypot functions, they design an active defense mechanism by integrating the decoy module, the coordinator module, and the captive module. Its performance is validated using real deployment and {tested} in a software-defined environment.
Based on software-defined network (SDN) and network function virtualization (NFV), Zarca \emph{et al}. \cite{9060972} design a virtual IoT honeypot network to realize flexible and programmable honeypot deployment and dynamic security policy enforcement for mitigated network attacks.
Wang \emph{et al}. \cite{9149314} propose a hybrid IoT honeypot architecture for malware defense, which consists of a high-interactive component in real IoT devices and a low-interactive component in VMs with Telnet/SSH services.
For small-size and mobile UAVs, Daubert \emph{et al}. \cite{Daub2018HoneyDrone} develop a low/medium-interaction honeypot prototype system named HoneyDrone based on Raspberry Pi, where UAV-specific protocols are simulated in the honeypot prototype to cheat attackers.

One can observe that existing literature mainly focuses on system architecture and protocol design for honeypots in the IoT and UAV networks, whereas the collaborative honeypot-based UAV defense mechanisms are topics that are understudied. Given the widespread, advanced, and covert cyber attacks on UAV applications, it is necessary to deploy large-scale cooperative UAV defense by sharing trapped attack data in UAVs' honeypots.

\vspace{-2mm}
\subsection{Game Modeling for Honeypot-Based Defense}\label{subsec:RELATEDWORK2}
In the literature, various game-theoretical honeypot deception mechanisms have been proposed to enhance defense effectiveness.
Garg \emph{et al}. \cite{4267549} investigate the honeypot deployment problem using a strategic game-theoretical deception model between attackers and the defender (i.e., the honeynet administrator) under imperfect information, where {the} Bayesian equilibrium strategies {of participants} are analyzed.
Tian \emph{et al}. \cite{9431233} study a honeypot defense game against APT attacks under industrial IoT, where the stable strategies of attackers and defenders are analytically derived under bounded rationality.
Wang \emph{et al}. \cite{7857804} design a honeypot architecture to capture DDoS traffic on smart meters and devise a Bayesian game-theoretical model to model the interactions between DDoS attackers and defenders in smart grids.
La \emph{et al}. \cite{7442780} propose a Bayesian game-based deception model in honeypot networks containing an attacker and a defender, where the attacker can deceive the defender by exhibiting various behavior patterns ranging from suspicious to seemingly normal under incomplete information.
Tian \emph{et al}. \cite{9397770} present a contract game model to motivate small-scale electricity suppliers (SESs) equipped with honeypots to contribute local defense data with power retailers to reduce system defense costs.
Tsemogne \emph{et al}. \cite{9435800} design a two-player stochastic zero-sum game model to mitigate IoT botnet propagation to search for the optimal honeypot placement policy for the defender to deceive the attacker.
Wahab \emph{et al}. \cite{8675527} propose a repeated Bayesian Stackelberg game model to detect smart attackers in the clouds, where the attack patterns are learned from risky VMs using honeypots by support vector machine (SVM) methods.

However, the above works are mainly built atop high-interaction honeypots on real hosts or VMs, which are inapplicable to UAV networks with high mobility and limited resources. In addition, UAVs' multi-dimensional private information, different levels of information asymmetry, and the defender's budget constraints are ignored in previous works on game-based honeypot defenses. Table~\ref{Table1} summarizes the key differences between our work and related {researches}. %models and incentive design

\section{System Model}\label{sec:SYSTEMMODEL}
This section introduces the system model, consisting of the network model, UAV mobility model, channel model, and threat model. {Table \ref{notationtable} summarizes the notations used in the remaining of this paper.}

\subsection{Network Model}\label{subsec:networkmodel}
Fig.~\ref{fig:model} depicts a typical honeypot-based collaborative defense scenario in a UAV network, consisting of one GCS (denoted as $G$) and $I$ flying UAVs. A group of UAVs (denoted as $\mathcal{I} = \{1,\cdots,I\}$) mounted with rich sensors are dispatched to a specific task area for immediate mission execution (e.g., power lines inspection). UAVs can exchange {flight} information for collision avoidance via air-to-air (A2A) links. %using on-board communication modules.
Besides, each UAV is equipped with a low/medium-interaction honeypot system to allow emulation, recording, and analysis of its captured malicious activities to mitigate cyber attacks. Let $S_i$ denote UAV $i$'s private valid defense data (VDD) volume, which means the data size of unknown attack interaction logs gathered by the UAV honeypot \cite{Daub2018HoneyDrone}. UAVs are distinguished by their 2D private information: the marginal VDD cost and the communication delay.
Let $\mathcal{J} = \{1,\cdots,J\}$ be the set of UAV's types. We refer to a UAV with $\theta_j \triangleq (C_j,T_j)$ as a type-$j$ UAV. Here, $C_j$ means the unit cost for VDD generation, VDD transmission, and privacy loss of type-$j$ UAV. $T_j$ is the communication delay of type-$j$ UAV in transmitting VDD amount $S_j$ (in bytes) to the GCS.

The GCS, as the coordinator of the UAV network, can communicate with UAVs via air-to-ground (A2G) links, perform UAV control (e.g., task assignment and trajectory planning), and carry out task data processing as well as security provisioning.
Traditionally, the GCS obtains defense data through external security service providers. In our scenario, the GCS additionally obtains defense data from UAVs, which have deployed the honeypot, for quicker attack recognition and better situational awareness.
To motivate UAVs' cooperation, the GCS offers a series of contracts $\Phi=\{T_{\max},\{\Phi_j\}_{j\in \mathcal{J}}\}$ including the maximum communication delay $T_{\max}$ (for all UAV's types) and $J$ contract bundles $\{\Phi_j\}_{j\in \mathcal{J}}=\{S_j,R_j\}_{j\in \mathcal{J}}$ (one for each type). Here, $S_j$ and $R_j$ are the contributed VDD size and contractual reward (i.e., compensation\footnotemark[2]) of each type-$j$ UAV, respectively.\footnotetext[2]{{The compensation form can be monetary payments or network tokens, and its detailed form depends on the specific UAV applications.}} For any UAV fails to deliver its VDD within $T_{\max}$, the GCS offers a zero-payment contract. It is assumed that the type of each UAV remains unchanged in the entire honeypot defense process.

\begin{table}%\scriptsize
\setlength{\abovecaptionskip}{-0.0cm}
{
\caption{Summary of Notations}
\label{notationtable}\centering
\begin{tabular}{|c|l|}
\hline
\textbf{Notation} & \textbf{Description} \\  \hline %\hline
$\mathcal{T}$&Set of time slots. \\
$\mathcal{I} \!=\! \{1,\cdots,I\}$&Set of honeypot-mounted UAVs in an investigated area.\\
$\mathcal{J} \!=\! \{1,\cdots,J\}$&Set of UAV's types in collaborative honeypot defense.\\
${N_j}$&Number of type-$j$ UAVs.\\
$G$&The GCS for coordinating the UAV network.\\
$C_j$&Unit cost of honeypot defense of type-$j$ UAV.\\
$T_j$&Delay of type-$j$ UAV in VDD transmission to the GCS.\\
$\theta_j \!=\! (C_j,T_j)$&2D type of UAV (VDD cost \& communication delay).\\
$T_{\max}$&Maximum communication delay for UAVs.\\
$S_j$&Contractual VDD size of type-$j$ UAV.\\
$S_{\max}$&Maximum contractual VDD size for a UAV.\\
$R_j$&Contractual reward of type-$j$ UAV.\\
$R_{\max}$&Maximum contractual reward to a UAV.\\
%$\Phi\!=\!\{T_{\max},\{\Phi_j\}\}$&\\
$\Phi_j \!=\! \{S_j,R_j\}$&$j$-th contract bundle (VDD size \& reward).\\
$\Omega$&Overall budget of the GCS in contract design.\\
$[{x_i}(t),{y_i}(t)]$& Instantaneous horizontal coordinate of UAV $i$.\\
${z}_i(t)$& Instantaneous altitude of UAV $i$.\\
$\Delta_t$& Length of each time slot.\\
${V_i}(t),V_{\max }^i$& Flying velocity and maximum velocity of UAV $i$.\\
$\Upsilon_{i,k}^{\mathrm{A2A}}(t)$& A2A channel path loss between UAV $i$ and UAV $k$.\\
$\Upsilon_{i,G}^{\mathrm{A2G}}(t)$& A2G channel path loss between UAV $i$ and the GCS.\\
$d_{i,k}(t)$& 3D Euclidean distance between UAV $i$ and UAV $k$.\\
$\gamma _{i,k},{\gamma _{i,G}}$& Available data rate from UAV $i$ to UAV $k$/the GCS.\\
$P_{i}^{\text{Tr}}(t)$& UAV $i$'s transmit power at time slot $t$.\\
${B^{\mathrm{A2A}}},{B^{\mathrm{A2G}}}$& A2A/A2G channel bandwidth.\\
${\Pr}_{\mathrm{{LoS}}},{\Pr}_{\mathrm{{NLoS}}}$& LoS/NLoS probability of A2G communications.\\
$\delta_{th}$& SNR threshold for UAVs to select A2A or A2G mode.\\
$\mathcal{U}_{j}\left( \Phi_j \right)$& Utility of type-$j$ UAV that chooses $j$-th contract item.\\
$\mathcal{U}_{G}( \Phi )$& Utility of the GCS in the designed contracts.\\
$\mathcal{S}( \Phi )$& Social surplus of all participating UAVs and the GCS.\\
${\bf{W}}^t \!=\! {\bf{S}}^{t-1}$& System state vector observed by the GCS.\\
$R_j^t$& Discrete reward action of type-$j$ UAV at time slot $t$.\\
$\mathcal{{Q}}\left({\bf{W}}^t, {\bf{R}}^t\right)$& Q-function of the GCS.\\
$\pi\left({\bf{W}}^t, {\bf{R}}^t\right)$& Mixed-strategy table of the GCS in PHC.\\
$\tilde{W}_j^t \!=\! {R}_{j}^{t-1}$& System state observed by type-$j$ UAV at time slot $t$.\\
$S_j^t$& Discrete VDD size action of type-$j$ UAV at time slot $t$.\\
$\tilde{\mathcal{{Q}}}\left(\tilde{W}_j^t, S_j^t\right)$& Q-function of type-$j$ UAV.\\
$\tilde{\pi}\left(\tilde{W}_{j}^t, S_{j}^t\right)$& Mixed-strategy table of type-$j$ UAV in PHC.\\
$\kappa_1,\kappa_2$& Learning rates in updating the Q-function.\\
$\varphi_1,\varphi_2$& Discount factors in updating the Q-function.\\
$\rho_1,\rho_2$& Greedy factors in updating the mixed-strategy table.\\
\hline
\end{tabular}}
\end{table}

\subsection{UAV Mobility Model}\label{subsec:mobilitymodel}
Based on \cite{9696188}, the total time horizon is evenly divided into $T$ time slots with time length $\Delta_t$. When $\Delta_t$ is sufficiently small, the instant location of UAV $i$ in each time slot can be approximately fixed. According to the 3D Cartesian coordinate system, UAV $i$'s instant 3D location at {time slot $t$} is denoted as ${\bf{l}}_i(t) = [x_i(t),y_i(t),z_i(t)],\forall t \in \mathcal{T},\forall i \in \mathcal{I}$. Here, $[{x_i}(t),{y_i}(t)]$ is the instantaneous horizontal coordinate of UAV $i$ at time slot $t$. For each UAV $i$, its hover height ${z}_i(t)$ is fixed during executing each mission to ensure continuous flight and avoid frequent ascent/descent for minimized energy consumption \cite{8413129,10106022}. The trajectories of UAVs are predetermined and controlled by the GCS $G$, which satisfies:
\begin{numcases}{}
{{\bf{l}}_i} = \left\{ {{{\bf{l}}_i}\left( 1 \right), \cdots ,{{\bf{l}}_i}\left( t \right), \cdots ,{{\bf{l}}_i}\left(T\, \right)} \right\}, \label{eq:mobility0} \hfill \\
{{\bf{l}}_i}(t+1) = {V_i}(t)\cdot{{\bf{w}}_i}(t) + {{\bf{l}}_i}(t),1 \le t \le T-1,\label{eq:mobility1} \hfill \\
{\rm{s.t.}}\; || {{\bf{l}}_i}(t+1) - {{\bf{l}}_i}(t) || \le \Delta_t V_{\max }^i. \label{eq:mobility2}
\end{numcases}
In Eq. (\ref{eq:mobility0}), ${{\bf{l}}_i}\left( {{1}} \right)$ and ${{\bf{l}}_i}\left( T \right)$ are the preset starting and ending locations of UAV $i$ in the working area, respectively.
In Eq. (\ref{eq:mobility1}), ${V_i}(t)$ and ${\bf{w}}_i(t)$ are the flying velocity and trajectory direction of UAV $i$ at {time slot $t$}, respectively. In Eq. (\ref{eq:mobility2}), $V_{\max }^i$ is UAV $i$'s maximum velocity.

\subsection{Channel Model}\label{subsec:channelmodel}
\emph{1) A2A Channel Model.} The A2A channel path loss (in dB) between UAVs $i$ and $k$ can be regarded as LoS-dominant and distance-dependent \cite{9115898}, i.e., $\Upsilon_{i,k}^{\mathrm{A2A}}(t) = \left(d_{i,k}(t)\right)^{ - \iota }$, where $\iota$ means the path loss exponent and $d_{i,k}(t)$ is the 3D Euclidean distance between two UAVs $i$ and $k$. {As the communications between UAVs are usually LoS, the effect of self-interference in UAV communications is ignored.}
Let ${B^{\mathrm{A2A}}}$ denote the A2A channel bandwidth.
At time slot $t$, based on the Shannon bound, the available data rate from UAV $i$ to UAV $k$ is
\begin{align}\label{eq:comm6}
\gamma _{i,k}(t) = {B^{\mathrm{A2A}}}{\log _2}\left( {1 + \frac{P_{i}^{\text{Tr}}(t)\Upsilon_{i,k}^{\mathrm{A2A}}(t)}{ \sum_{l\in \mathcal{I},l \ne i} {P_{l}^{\text{Tr}}(t)\Upsilon_{l,k}^{\mathrm{A2A}}(t)}+ {\varphi^2}} } \right),
\end{align}
where $P_{i}^{\text{Tr}}(t)$ is UAV $i$'s transmit power at time slot $t$. $\sum_{l\in \mathcal{I},l \ne i} {P_{l}^{\text{Tr}}(t)\Upsilon_{l,k}^{\mathrm{A2A}}(t)}$ represents the sum of interferences from other UAVs to UAV $k$ at time slot $t$. ${\varphi^2}$ is the power of the additive white Gaussian noise.

\emph{2) A2G Channel Model.} For A2G/G2A communications, the average pathloss (in dB) between UAV $i$ and the GCS follows the large-scale channel fading model depending on the occurrence chances of LoS and non-LoS (NLoS) links \cite{8624565}, i.e.,
\begin{align}\label{eq:A2Gpathloss}
\Upsilon_{i,G}^{\mathrm{A2G}}(t)= 20 &\log \big({{4 \pi d_{i,G} \phi_{c}}/{c}}\big) + {\Pr}_{\mathrm{{LoS}}}(t) \kappa_{\mathrm{{LoS}}} \nonumber \\ &+ {\Pr}_{\mathrm{{NLoS}}}(t) \kappa_{\mathrm{{NLoS}}},
\end{align}
where $\kappa_{\mathrm{{LoS}}}$ and $\kappa_{\mathrm{{NLoS}}}$ are additional attenuation factors of free space pathloss for LoS and NLoS links, respectively. $\phi_{c}$ is the carrier frequency, $c$ means the speed of light, and $d_{i,G}$ is the horizontal distance between UAV $i$ and the GCS. The LoS probability ${\Pr}_{\mathrm{{LoS}}}(t)$ is a modified logistic function of the elevation angle $\theta_{i,G}(t)=\arctan(\frac{z_i(t)-h_G}{d_{i,G}})$ \cite{8663615}, i.e.,
\begin{align}\label{eq:los}
{\Pr}_{\mathrm{{LoS}}}(t)= \left[ 1+ \iota_1\exp(-\iota_2(\theta_{i,G}(t)-\iota_1))\right]^{-1}.
\end{align}
Here, $\iota_1$ and $\iota_2$ are environment-related variables, $h_G$ is the height of GCS, and ${\Pr}_{\mathrm{{NLoS}}}(t)=1-{\Pr}_{\mathrm{{LoS}}}(t)$.

For A2G/G2A data transmissions, based on works \cite{8758340,8663615,8624565}, it is assumed that each UAV is allocated a dedicated sub-channel with orthogonal resource blocks for uplink transmission. Thereby, there exists no mutual interference between UAVs. Let ${B^{\mathrm{A2A}}}$ denote the uplink A2G channel bandwidth. At time slot $t$, the available uplink data rate from UAV $i$ to the GCS is
\begin{align}\label{eq:comm5}
{\gamma _{i,G}}(t) = {B^{\mathrm{A2G}}}{\log _2}\left( {1 + \frac{P_{i}^{\text{Tr}}(t)\cdot 10^{-\Upsilon_{i,G}^{\mathrm{A2G}}(t)/10}}{{\varphi^2}}} \right).
\end{align}

\subsection{Threat Model}\label{subsec:threatmodel}
In the cooperative UAV defense services based on honeypot data sharing, the following two threats that may deteriorate system efficiency and defense performance are considered.
\begin{itemize}
   \item \textbf{Selfish UAVs:} {UAVs'} participation is the key to the success of collaborative UAV defense services. However, as the deployment of honeypots and the VDD data transmission operations can consume additional battery power of resource-limited UAVs, UAV owners (as rational and selfish individuals) may be reluctant to participate in the joint defense process \cite{9036917}. Thereby, the overall defense performance can be reduced as it lacks enough participants.
   \item \textbf{Free-riding UAVs:} Self-interested UAVs may launch free-riding attacks to gain an unfair advantage without contributing to the joint defense process, thereby inhibiting the enthusiasm and willingness of other honest UAVs \cite{9159929}. For example, free-riding UAVs may enjoy joint defense services by sharing redundant and outdated honeypot data to cheat more rewards {from the GCS}.
\end{itemize}

\section{Honeypot Game Theoretical Cooperative Defense Framework}\label{sec:GAME}
In this section, we first formulate the UAV honeypot game for active cooperative defense among UAVs. Then, we define the equilibrium and design goals of the UAV honeypot game.
\subsection{{One-Shot} UAV Honeypot Game Formulation}\label{subsec:Honeypotmodel}
\begin{definition}[UAV Honeypot Game]
During honeypot data sharing process, the interactions between UAVs and the GCS can be formulated as a honeypot game $\mathcal{G}=\{\{\mathcal{J},G\},\{T_{\max},\{S_j,R_j\}_{j\in \mathcal{J}}\},\{\{\mathcal{U}_j\}_{j\in \mathcal{J}},\mathcal{U}_G\}\}$, which includes the following main components:
\begin{itemize}
  \item \emph{Players.} The players in game $\mathcal{G}$ are (i) UAVs with diverse VDD-delay types in {the set} $\mathcal{J}$ and (ii) the GCS $G$. {In the game, UAVs are featured with 2D types: the VDD cost $C_j$ and the communication delay $T_j$, which are private information.}
  \item \emph{Strategies.} The strategy of the GCS is to determine the maximum communication latency $T_{\max}$ and design a set of feasible VDD-reward contracts $\{S_j,R_j\}_{j\in \mathcal{J}}$ to optimize its overall payoff. The strategy of each {participating} UAV is to select an optimal contract item {from the contract menu $\Phi$} for {its} maximized payoff. %to specify the relation between VDD volumes and rewards
  \item \emph{Payoffs.} The payoffs (or utilities) of each type-$j$ UAV and the GCS are denoted as $\mathcal{U}_j$ and $\mathcal{U}_G$, respectively.
\end{itemize}

{Note that the game $\mathcal{G}$ is \emph{one-shot}, i.e., the game is played once between the GCS and each honeypot-mounted UAV.}
\end{definition}

\underline{Utility of UAV.} The utility of type-$j$ UAV that chooses the contract item $\Phi_j=\{S_j,R_j\}$ is the revenue minuses its cost:
\begin{align}\label{eq:utility-UAV}
\mathcal{U}_{j}\left( \Phi_j \right) \!=\! \left\{ \begin{array}{ll}
	R_j\!-\!{C_j^1}S_j\!-\!{C_j^2}S_j\!-\!{C_0}, &\mathrm{if}\, T_j\le T_{\max};\\
	-{C_j^1}S_j-{C_j^2}S_j-{C_0},& \mathrm{if}\, T_j> T_{\max}.
\end{array} \right.
\end{align}

In (\ref{eq:utility-UAV}), ${C_j^1}$ is the unit cost of VDD creation and transmission {of type-$j$ UAV}, which is related to UAV's honeypot and communication capabilities. ${C_j^2}$ is the unit privacy cost of VDD sharing {of type-$j$ UAV}. Both ${C_j^1}$ and ${C_j^2}$ are UAV's private information. Here, ${C_j}={C_j^1}+{C_j^2}$. $C_0$ is UAV's honeypot deployment cost. {$T_j> T_{\max}$ means that the type-$j$ UAV can transmit its VDD with size $S_j$ within $T_{\max}$.}

To improve communication efficiency, an \emph{A2A/A2G mode selection method} is designed. Specifically, if UAV $j$ experiences a high signal-to-noise ratio (SNR) for the A2G link, it directly uploads its encrypted VDD via A2G mode. Otherwise, it alternatively delivers data to a neighboring UAV $j'$ operating in A2G mode and relays to the GCS. {Let $\delta_{th}$ be the SNR threshold to distinguish the UAVs that work in A2A and A2G modes.} Let $\alpha_j=\{0,1\}$ be a binary variable, where $\alpha_j=1$ means it works on A2G mode, otherwise $\alpha_j=0$. We have
\begin{align}\label{eq:latency}
T _{j}= {\alpha_j}\times\frac{S_j}{\gamma _{j,G}} + (1 - {\alpha_j})\times \Big(\frac{S_j}{\gamma_{j,j'}} + \frac{S_j}{\gamma _{j',G}}\Big),
\end{align}
where ${\gamma _{j,G}}$ and ${\gamma _{j',G}}$ are data rates between UAV $j$/$j'$ and GCS $G$ according to Eq. (\ref{eq:comm5}). ${\gamma_{j,j'}}$ is the data rate between UAVs $j$ and $j'$ according to Eq. (\ref{eq:comm6}).

\underline{Utility of GCS.} The utility of the GCS is the overall satisfaction of cooperative defense minuses its total payments:
\begin{align}\label{eq:utility-GCS}
\mathcal{U}_{G}( \Phi )= \sum\limits_{j \in \mathcal{J}} \varpi \frac{N_j}{T_j}\log \left(1 + \mathbbm{1}_{T_j\le T_{\max}} S_j \right) - \mathbbm{1}_{T_j\le T_{\max}} N_j R_j.
\end{align}

In (\ref{eq:utility-GCS}), the first term indicates the satisfaction related to UAV's VDD size and communication latency. %and reputation value.
Based on \cite{9035635,5738226}, we utilize the natural logarithmic function for satisfaction modelling. $\varpi$ is a positive satisfaction factor. ${N_j}$ is the number of type-$j$ UAVs, which satisfies $\sum_{j\in \mathcal{J}}{N_j}=I$.
$\mathbbm{1}_{T_j\le T_{\max}}$ is an indicator function, whose value equals to one if ${T_j\le T_{\max}}$ holds; otherwise its value is zero.

\underline{Social Surplus.} The social surplus of the UAV honeypot game is defined as the sum of the utilities of the GCS and all participating UAVs in collaborative honeypot defense, i.e.,
\begin{align}\label{eq:SocialSurplus}
&\mathcal{S}( \Phi )= \mathcal{U}_{G}( \Phi ) + \sum\limits_{j \in \mathcal{J}} {\mathbbm{1}_{T_j\le T_{\max}}\mathcal{U}_{j}\left( \Phi_j \right)}  \\
&\!=\!\sum\limits_{j \in \mathcal{J}} \varpi \frac{N_j}{T_j}\log \left(1 \!+\! \mathbbm{1}_{T_j\le T_{\max}} S_j \right) \!-\! \mathbbm{1}_{T_j\le T_{\max}} N_j\left( {C_j}{S_j}\!-\!C_0 \right).\nonumber
\end{align}

\subsection{Equilibrium and Design Goals of UAV Honeypot Game}\label{subsec:gameequilibrium}

The equilibrium strategy of the game $\mathcal{G}$ (i.e., the solution of the game) is to design the \emph{optimal contracts} for all types of UAVs, namely, $\Phi^*=\{T_{\max},\{S_j^*,R_j^*\}_{j\in \mathcal{J}}\}$, while enforcing \emph{budget feasibility}, \emph{contractual feasibility}, and \emph{contractual fairness}. Budget feasibility (BF) means that the GCS {can} only afford a constrained system budget (i.e., limited overall rewards) in each defense process in the honeypot game.

\begin{definition}[Budget Feasibility (BF)]
A contract is budget feasible, if the total reward for all participating UAVs does not exceed the overall budget $\Omega$, i.e.,
\begin{align}\label{eq:4-BF}
\mathbbm{1}_{T_j\le T_{\max}} N_j R_j \le \Omega.
\end{align}
\end{definition}

Apart from {the} BF {property}, contractual feasibility and optimality are basic goals of incentive mechanism design, which are formally defined as follows.

\begin{definition}[Contractual Feasibility]
A contract {$\Phi$} is feasible if each type of UAV has the greatest and non-negative utility when faithfully adopting the contract item designed for its type.
\end{definition}

\begin{definition}[Contractual Optimality]
Among all feasible contracts, a contract {$\Phi$} is optimal if it maximizes the utility of the contract designer (i.e., the GCS).
\end{definition}

According to the revelation principle \cite{Contract2005Bolton}, a contract satisfying the contractual feasibility is equivalent to that the individual rationality (IR) and incentive compatibility (IC) constraints are satisfied simultaneously for all types of UAVs. The IR and IC constraints are formally defined as follows{.}

\begin{definition}[Individual Rationality (IR)]
If and only if each type-$j$ UAV can obtain non-negative utility when faithfully adopting the contract item $\Phi_{j}=\{S_{j},R_{j}\}$ designed for its type, then the contract {$\Phi$} satisfies the IR constraint. Mathematically,
\begin{align}\label{eq:4-1-IR}
\mathcal{U}_{j}\left( \Phi_j \right) \geq 0,\forall j \in \mathcal{J}.
\end{align}
\end{definition}

\begin{definition}[Incentive Compatibility (IC)]
If and only if each type-$j$ UAV prefers to faithfully adopt the contract item $\Phi_{j}=\{S_{j},R_{j}\}$ designed for its type rather than other contract items, then the contract {$\Phi$} satisfies the IC constraint. Mathematically,
\begin{align}\label{eq:4-1-IC}
   \mathcal{U}_{j}\left( \Phi_j \right) \geq \mathcal{U}_{j}\left( \Phi _{j{'}} \right), \forall j, j{'} \in \mathcal{J}, j\ne j{'}.
\end{align}
\end{definition}

In addition to optimality, fairness is another desirable target of incentive mechanism design. Based on the literature \cite{9296813}, the definitions of participation fairness and reward fairness are first introduced. Then, the contractual fairness is defined based on these two aspects.

\begin{definition}[Participation Fairness]
Participation fairness is satisfied if any rational and selfish UAV honestly follows the contractual procedure. Namely, they have no incentive to withdraw from the collaborative honeypot data sharing process and report false individual types to demand more {compensations}.
\end{definition}

\begin{definition}[Reward Fairness]
If 1) higher rewards are given to participating UAVs that contribute more VDD in collaborative defense, and 2) no reward is given to non-participating UAVs, then {the} reward fairness is satisfied.
\end{definition}

\begin{definition}[Contractual Fairness]
If both participation fairness and reward fairness are satisfied, then the contract is said to be fair.
\end{definition}

In this paper, we design the optimal contracts for UAVs to solve the honeypot game under the following three levels of information asymmetry.
\begin{itemize}
  \item \emph{Complete information scenario (benchmark)}. In this ideal situation, there exists no information asymmetry and the GCS knows the private type information of each UAV.
  \item \emph{Partial information asymmetry scenario}. The GCS {only knows} the distribution of UAV's types (i.e., $N_j/I,\forall j \in \mathcal{J}$) and the total number of UAVs (i.e., $I$), but is unaware of which UAV belongs to which type. Note that the distribution of UAV's types can be obtained in various manners, e.g., making a survey questionnaire.
  \item \emph{Complete information asymmetry scenario}. {The GCS does not have any knowledge of UAVs' private types in the honeypot game even for its distribution information.} The GCS only knows the total number of UAVs and the total number of UAV's types. {Besides, after multiple honeypot data-sharing interactions, the GCS is aware of the historical strategy information of participating UAVs, while any participating UAV knows the GCS's historical strategy information to itself.}
\end{itemize}

\section{Optimal Contract Design in Complete Information}\label{sec:CompleteSolution}
In the complete information scenario, the contract designer (i.e., the GCS) knows the private type of each UAV. Thereby, it can check whether UAVs faithfully adopt the contract items designed for their types. Correspondingly, the GCS only needs to ensure that all types of UAVs can obtain non-negative {utilities} in the honeypot game. Therefore, the contract feasibility constraint is equivalent to the IR constraint.

\subsection{Optimization Problem in Complete Information}
\begin{problem}[GCS's optimization problem under complete information scenario]
\begin{align}\label{eq:Comp-optproblem1}
\begin{gathered}
   \mathop {\max }\nolimits_{\Phi} \mathcal{U}_{G}\left( \Phi \right) \hfill \\
   ~\mathrm{s.t.}\;\left\{ \begin{gathered}
   0\le S_{j}\le S_{\max},\forall j \in \mathcal{J}, \hfill \\
   {\text{BF constraint (\ref{eq:4-BF})}},\forall j \in \mathcal{J}, \hfill \\
   {\text{IR constraint (\ref{eq:4-1-IR})}},\forall j \in \mathcal{J}. \hfill \\
\end{gathered} \right. \hfill \\
\end{gathered}
\end{align}

\emph{Remark.}
The first constraint means that the amount of VDD contributed by each type of UAV is constrained by the upper bound $S_{\max}$ and the lower bound $0$. The second one is the BF constraint, and the third one is the IR constraint.
\end{problem}

\subsection{Optimal Contract in Complete Information}
{Due to the existence of the indicator function, the objective function $\mathcal{U}_{G}\left( \Phi \right)$ in Eq. (\ref{eq:utility-GCS}) in the Problem~1 is non-convex. We classify the UAVs into \emph{participating ones} and \emph{non-participating ones}.} To simplify the expression, let $\mathcal{J}'$ be the set of {UAV's} types that {satisfies} $\mathbbm{1}_{T_j\le T_{\max}}=1$, i.e., $\mathcal{J}'=\{j|{T_j\le T_{\max}}\}$. Then we reindex {UAV's} types in $\mathcal{J}'$ in descending order of the marginal VDD cost, i.e., $C_1>C_2>\cdots>C_{J'}$, where $J'=|\mathcal{J}'|$. In other words, UAVs with type $j \notin \mathcal{J}'$ cannot participate in the collaborative honeypot defense and will receive no payment.

Next, we solve the Problem 1 in two steps. First, for any given VDD size, Lemma \ref{lemma2-1} gives the optimal reward policy for the GCS. Second, by substituting the optimal reward strategy into the GCS's utility function, Theorem \ref{theorem2-2} proves the optimal {contractual} VDD size strategy.

\begin{lemma}\label{lemma2-1}
For any VDD data size $S_{j}\in[0, S_{\max}]$, the optimal reward strategy {of} the GCS is:
\begin{align}\label{eq:Complete-OptimalPrice}
     R_{j}^{*}\left(S_{j}\right) =\left\{ \begin{array}{ll}
     0,&\forall j \notin \mathcal{J}';\\
     {C_j} S_{j} + C_0,&\forall j \in \mathcal{J}'.\\
     \end{array} \right.
     \end{align}
\end{lemma}

\begin{IEEEproof}
Please refer to Appendix A.
\end{IEEEproof}

\begin{lemma}\label{lemma2-1-2}
The BF constraint in (\ref{eq:4-BF}) can be simplified as:
\begin{align}\label{eq:BF-simp}
\sum\nolimits_{j=1}^{J'} {N_j R_j} = \Omega.
\end{align}
\end{lemma}

\begin{IEEEproof}
Please refer to Appendix B.
\end{IEEEproof}

\emph{Remark.}
Lemmas~\ref{lemma2-1} and \ref{lemma2-1-2} mean that: under the complete information scenario, for non-participating UAVs, the GCS will provide a zero-payment contract; for participating UAVs, the GCS will design optimal contract items by exhausting the available budget (i.e., $\Omega$) such that all participating UAVs will receive zero utility.

For any participating UAV ($\forall j \in \mathcal{J}'$), by substituting $R_{j}^{*}={C_j} S_{j} + C_0$ {derived from (\ref{eq:Complete-OptimalPrice})} into $\mathcal{U}_{G}( \Phi )$ in (\ref{eq:utility-GCS}), the utility function of the GCS can be rewritten as a function of $S_{j}$:
\begin{align}\label{optimalOpUtilityCIS}
\mathcal{U}_{G}(S_j)\!=\! \sum\nolimits_{j \in \mathcal{J}'} \frac{\varpi N_j}{T_j}\log \left(1 + S_j \right) \!-\! N_j({C_j} S_{j} \!+\! C_0).
\end{align}
Based on Lemmas~\ref{lemma2-1} and \ref{lemma2-1-2}, the optimization problem {in} (\ref{eq:Comp-optproblem1}) can be equivalently formulated as below.

\emph{\textbf{Problem} 1-1 (Simplified Problem 1 with reduced BF and IR constraints):}
\begin{align}\label{eq:Comp-optproblem1-2}
\begin{gathered}
   \mathop {\max }\nolimits_{\Phi} \mathcal{U}_{G}\left( S_j \right) \hfill \\
   ~\mathrm{s.t.}\;\left\{ \begin{gathered}
   0\le S_{j}\le S_{\max},\forall j \in \mathcal{J}', \hfill \\
   \sum\nolimits_{j\in \mathcal{J}'} {N_j R_j} = \Omega, \hfill \\
   R_{j}= {C_j} S_{j} + C_0,\forall j \in \mathcal{J}'. \hfill \\
\end{gathered} \right. \hfill \\
\end{gathered}
\end{align}

The above Problem 1-1 can be solved via Lagrange analysis with KKT conditions, {and} its Lagrangian function is:
\begin{align}\label{eq:Lagrangian1}
{\mathscr{L}}(S_j,\lambda_1) = \mathcal{U}_{G}(S_j) + \lambda_1 \Big(\sum\nolimits_{j=1}^{J'} {N_j \left({C_j} S_{j} + C_0 \right)} - \Omega\Big) \nonumber \\
\!=\!\sum\limits_{j =1}^{J'} \bigg[{\frac{\varpi N_j}{T_j}\log \left(1 \!+\! S_j \right) + (\lambda_1 \!-\! 1)N_j({C_j} S_{j} \!+\! C_0)} \bigg] - \lambda_1 \Omega,
\end{align}
where $\lambda_1$ denotes the Lagrange multiplier.

The following Theorem \ref{theorem2-2} further deduces the optimal contractual VDD size for each type of UAV.

\begin{theorem}\label{theorem2-2}
Under the complete information scenario, the {contractual VDD size and contractual reward for each type of UAV} in the optimal contract {$\Phi^*=\{T_{\max},\{S_j^*,R_j^*\}_{j\in \mathcal{J}}\}$} are:
\begin{enumerate}
   \item $\forall j \notin \mathcal{J}'$, $S_{j}^{*}=R_{j}^{*}=0$.
   \item $\forall j \in \mathcal{J}'$, we have
     \begin{numcases}{}	
     S_{j}^{*}\!=\!\min\left\{S_{\max},\max\left\{\frac{\Lambda}{T_j C_j} -1 ,0\right\}\right\}, \label{eq:CIoptS} \hfill \\
     R_{j}^{*}= {C_j} S_{j}^{*} + C_0, \label{eq:CIoptR} \hfill
 \end{numcases}
where {$\Lambda$ is the abbreviation for}
\begin{align}\label{eq:Lambda}
\Lambda = \frac{\Omega + \sum\nolimits_{j=1}^{J'}{N_j {C_j}}- C_0\sum\nolimits_{j=1}^{J'}{N_j}}{\sum\nolimits_{j =1}^{J'}{\frac{N_j}{T_j}}}.
\end{align}
\end{enumerate}
\end{theorem}

\begin{IEEEproof}
Please refer to Appendix C.
\end{IEEEproof}

{\emph{Remark.}
For all non-participating UAVs, Theorem~1 shows that both the contractual VDD size and reward are zero.
For all participating UAVs, the optimal strategy on UAVs' contributed VDD size (i.e., $S_{j}^{*}$) is constrained by the upper bound $S_{\max}$ and the lower bound $0$. Meanwhile, under the complete information scenario, $S_{j}^{*}$ is determined by the UAV's type information (i.e., VDD cost ${C_j}$ and communication delay ${T_j}$), deployment cost $C_0$ of UAV honeypot, number of each type of UAVs ${N_j}$, and budget $\Omega$. Besides, the optimal reward strategy $R_{j}^{*}$ is a linear function of the corresponding VDD size in the contract.
}

\begin{corollary}\label{corollary0}
From (\ref{optimalOpUtilityCIS}) {and (\ref{eq:SocialSurplus})}, it can be deduced that in the complete information scenario, the GCS's optimal utility is equivalent to the optimal social surplus. Therefore, the optimal contract under the complete information scenario derived in Theorem~\ref{theorem2-2} is also a social optimal contract strategy.
\end{corollary}

\section{Optimal Contract Design in Partial Information Asymmetry}\label{sec:StaticSolution}
Unlike the complete information scenario, there usually exists information asymmetry between the GCS and UAVs in practical applications, where the optimization problem under incomplete information scenarios is formulated {in Sect.~\ref{subsec6-1}}.
In the partial information asymmetry scenario, the GCS only knows the total number of UAVs (i.e. $I$) and the private type distribution of UAVs (i.e., $p_j=\{\frac{N_j}{I}\}_{ \forall j \in \mathcal{J}}$).

\subsection{Optimization Problem in Incomplete Information}\label{subsec6-1}
\begin{problem}[GCS's optimization problem under incomplete information scenario]
\begin{align}\label{eq:optproblem1}
\begin{gathered}
  \mathop {\max }\nolimits_{\Phi}\, \mathcal{U}_{G}\left( \Phi \right)  \hfill \\
  ~\mathrm{s.t.}\;\left\{ \begin{gathered}
  0\le S_{j}\le S_{\max},\forall j \in \mathcal{J}, \hfill \\
  {\text{BF constraint (\ref{eq:4-BF})}},\forall j \in \mathcal{J}, \hfill \\
  {\text{IR constraint (\ref{eq:4-1-IR})}},\forall j \in \mathcal{J}, \hfill \\
  {\text{IC constraint (\ref{eq:4-1-IC})}},\forall j \in \mathcal{J}. \hfill \\
\end{gathered}  \right. \hfill \\
\end{gathered}
\end{align}

\emph{Remark.}
The first three constraints are the same as Problem~1, and the fourth constraint is the IC constraint {defined} in (\ref{eq:4-1-IC}). {According to Definitions 3, 5 and 6,} constraints (\ref{eq:4-1-IR}) and (\ref{eq:4-1-IC}) {jointly} enforce the contractual feasibility.
\end{problem}

Notably, there are $J^2$ IR and IC constraints in (\ref{eq:4-1-IR}) and (\ref{eq:4-1-IC}) in Problem~2, making it difficult to resolve Problem~2, particularly when $J$ is large. In what follows, IR and IC constraints are first transformed with reduced numbers using Lemma \ref{lemma2-3} and Theorem \ref{theorem2-4}. Then, given an arbitrary monotonic VDD size sequence $\mathbf{S}$, Theorem \ref{theorem2-5} gives the optimal reward policy $\mathbf{R}^*(\mathbf{S})$. Then, based on these two theorems, Problem~2 is transformed into the equivalent Problem~2-1 with reduced constraints, and Theorem \ref{theorem2-6} derives the optimal VDD size sequence $\widetilde{\mathbf{S}}^*$ for the relaxed form of the problem without the monotonicity constraint. Lastly, according to the rationale in Theorem \ref{theorem2-7}, an optimal dynamic allocation algorithm is designed in Algorithm~\ref{Algorithm2-1} to acquire the optimal VDD size strategy ${\mathbf{S}}^*$ and the optimal reward strategy $\mathbf{R}^*({\mathbf{S}}^*)$ under budget constraints.

\subsection{Optimal Contract in Partial Information Asymmetry}
\begin{lemma}\label{lemma2-3}
If IC constraints in (\ref{eq:4-1-IC}) hold for all UAV's types, then IR constraints in (\ref{eq:4-1-IR}) can be replaced by $\mathcal{U}_{1}( \Phi_1 ) \geq 0$.
\end{lemma}

\begin{IEEEproof}
As the IC constraint is satisfied for all UAV's types, we can obtain
\begin{align}\label{eq:IRreducedPIS}
R_j\!-\!{C_j}S_j\!-\!{C_0} \!\ge\! R_1\!-\!{C_j}S_1\!-\!{C_0} \!\ge\! R_1\!-\!{C_1}S_1\!-\!{C_0}.
\end{align}
{From (\ref{eq:IRreducedPIS}), we have $\mathcal{U}_{j}( \Phi_j ) \ge R_1\!-\!{C_1}S_1-{C_0} = \mathcal{U}_{1}( \Phi_1 )$. If the IR constraint holds for type-1 UAV (i.e., $\mathcal{U}_{1}( \Phi_1 ) \geq 0$), then we have $\mathcal{U}_{j}( \Phi_j ) \geq 0$, $\forall j \in \mathcal{J}'$. Lemma~\ref{lemma2-3} is proved.}
\end{IEEEproof}

{\emph{Remark.}} Lemma~\ref{lemma2-3} implies that if type-$1$ UAV satisfies the IR constraint, then all types of UAVs satisfy the IR constraint.

{Based on Lemma~\ref{lemma2-3},} we further characterize the feasibility of the contract in the following {theorem}.

\begin{theorem}\label{theorem2-4}
A contract $\Phi=\{T_{\max},\{\Phi_j\}_{j\in \mathcal{J}}\}$ is feasible if and only if the following conditions {hold}:
\begin{enumerate}
  \item $\forall j \notin \mathcal{J}'$, $S_{j}=R_{j}=0$.
  \item $\forall j \in \mathcal{J}'$, we have
  \begin{numcases}{}	
     0\le S_1 \le \cdots \le S_{J'} \&\, 0\le R_1 \le \cdots \le R_{J'},~~~~\ \label{eq:theo1cons1} \hfill \\
     R_1-{C_1}S_1-{C_0}\ge 0,  \label{eq:theo1cons2} \hfill \\
     \begin{gathered}
     C_{j}\left( S_j - S_{j-1} \right) \le R_{j} - R_{j-1}~~~~~~~~~~~~~~~~~~~ \\~~~~~~~~\le C_{j-1}\left( S_{j}-S_{j-1} \right),\ j = 2,\cdots,{J'}. \label{eq:theo1cons3} \hfill
     \end{gathered}
 \end{numcases}
\end{enumerate}
\end{theorem}

\begin{IEEEproof}
Please refer to Appendix D.
\end{IEEEproof}

\emph{Remark.}
For any UAV with type $j \notin \mathcal{J}'$, Theorem~2 shows that the required contractual VDD size and reward are zero.
For the case $j \in \mathcal{J}'$, constraints (\ref{eq:theo1cons1}) and (\ref{eq:theo1cons3}) correspond to IC constraints, while constraint (\ref{eq:theo1cons2}) corresponds to IR constraints.
Constraint (\ref{eq:theo1cons1}) means that the GCS should demand more VDD from UAVs with smaller marginal costs and offer more rewards to them.
Constraint (\ref{eq:theo1cons2}) indicates that if the UAV with the highest marginal cost satisfies the IR constraint, then all types of UAVs meet IR constraints.
Constraint (\ref{eq:theo1cons3}) implies that if type-$j$ and type-$(j\!-\!1)$ UAVs satisfy the IC constraint, then type-$j$ UAV and any other type of UAV also satisfy the IC constraint.

\begin{corollary}\label{corollary1}
For any feasible contract item $\{S_j,R_j\}_{j\in \mathcal{J}'}$, the utilities of different types of UAVs satisfy:
\begin{align}\label{eq:UAVutilityorder}
\mathcal{U}_{1}\left( \Phi_{1} \right) < \cdots < \mathcal{U}_{j}\left( \Phi_{j} \right) < \cdots < \mathcal {U}_{J{'}}\left( \Phi_{J{'}} \right), \forall j \in \mathcal{J}'.
\end{align}
\end{corollary}

\begin{IEEEproof}
According to Theorem \ref{theorem2-4}, the UAV that requires more rewards should provide more VDD data, namely, $R_j \ge R_{k}$ and $S_j \ge S_{k}$ meet simultaneously. If $C_j < C_{k}$, we have
\begin{align}
\mathcal{U}_{j}( \Phi_{j} ) &= R_j - {C_j}{S_j} - C_0 \nonumber \\
&\ge R_k - {C_j}{S_k} - C_0~~\mathrm{(IC)} \nonumber \\
&> R_k - {C_k}{S_k} - C_0 = \mathcal{U}_{k}\left( \Phi_{k} \right).
\end{align}
It can be concluded that when $C_{k} > C_j$, we have $\mathcal{U}_{k}( \Phi_{k} ) < \mathcal{U}_{j}( \Phi_{j} )$. Since $C_1>C_2>\cdots>C_{J{'}}$, we have $\mathcal{U}_{1}( \Phi_{1} ) < \cdots < \mathcal{U}_{ j}( \Phi_{j} ) < \cdots < \mathcal{U}_{J{'}}( \Phi_{J{'}} )$, $\forall j \in \mathcal{J}'$.
\end{IEEEproof}

In the following Theorem~3, we derive the optimal reward strategy $\mathbf{R}^*(\mathbf{S})$. %given any monotonic VDD size sequence $\mathbf{S}$.

\begin{theorem}\label{theorem2-5}
Given any VDD size sequence $\mathbf{S} = \{S_{j}\}_{j \in \mathcal{J}'}$ meeting $0\le S_{1} \le \cdots \le S_{J'} \le S_{\max}$, the unique optimal reward strategy $\mathbf{R}^*=\{R_{j}^*\}_{j \in \mathcal{J}'}$ is attained by:
\begin{enumerate}
  \item $\forall j \notin \mathcal{J}'$, $R_{j}^*(\mathbf{S})=0$.
  \item $\forall j \in \mathcal{J}'$, we have
  \begin{align}\label{eq:optreward1}
    R_{j}^{*}\left( \mathbf{S} \right) =\left\{ \begin{array}{l}
        {C_j} S_{j} + C_0,\ ~~~~j=1;\\
    	R_{j-1}^{*}\left( \mathbf{S} \right) +C_j\left( S_{j}-S_{j-1} \right) ,\\ ~~~~~~~~~~~~~~~~~~ j=2,...,J'.\\	
    \end{array} \right.
  \end{align}
\end{enumerate}
\end{theorem}

\begin{IEEEproof}
Please refer to Appendix E.
\end{IEEEproof}

\emph{Remark.}
Theorem 3 shows that the optimal reward positively correlates with UAV's shared VDD size, thereby ensuring reward fairness. The BF constraint in (\ref{eq:4-BF}) can be simplified as $\sum\nolimits_{j=1}^{J'} {N_j R_j} = \Omega$. The proof is similar to that of Lemma~\ref{lemma2-1-2}.

According to Theorems 2--3, the original Problem 2 can be rewritten into the following simplified form.

\emph{\textbf{Problem} 2-1 (Simplified Problem 2 with reduced BF, IR, and IC constraints):}
\begin{align}\label{eq:optproblem2}
\begin{gathered}
  ~~~\mathop {\max }\nolimits_{\Phi}\, \mathcal{U}_{G}\left( \Phi \right)  \hfill \\
  \mathrm{s.t.}\left\{ \begin{gathered}
  \mathrm{C1:}\ 0\le S_1 \le \cdots \le S_{J{'}} \le S_{\max}, \hfill \\
  \mathrm{C2:}\ R_1-{C_1}S_1-{C_0}= 0, \hfill \\
  \mathrm{C3:}\ R_{j} \!-\! {C_j}S_{j} \!=\! R_{j-1} \!-\! {C_j}S_{j-1},\forall j\!=\!2,\!\cdots\!,J', \hfill \\
  \mathrm{C4:}\ \sum\nolimits_{j \in \mathcal{J}'} {N_j R_j} = \Omega. \hfill \\
\end{gathered}  \right. \hfill \\
\end{gathered}
\end{align}
%\end{problem}

Besides, {for any UAV with type $j \in \mathcal{J}'$,} the optimal reward strategy in (\ref{eq:optreward1}) can be reformulated by iteration as follows:
\begin{align}\label{eq:4-1-OptimalPrice}
R_{j}^{*}\left( \mathbf{S} \right) =\left\{ \begin{array}{l}
	{C_j}{S_{j}} + \sum\nolimits_{k=1}^{j-1}{\left( C_{k}-C_{k+1}\right){S_{k}}}+ C_0,\\ ~~~~~~~~~~~~~~~~~~~ j=2,\cdots,J';\\
	{C_j}{S_{j}}+ C_0,\ ~~~~j=1.\\
\end{array} \right.
\end{align}

\begin{theorem}\label{theorem2-6}
Under partial information asymmetry, the optimal contractual VDD size strategy to solve the {relaxed} Problem~2-1 without constraint C1 is attained as:
\begin{align}\label{eq:OptimalSize1}
S_{j}^*=\min\left\{S_{\max},\max\left\{\frac{N_j }{A_j T_j}\cdot \mathfrak{R} -1 ,0\right\}\right\},
\end{align}
where $\mathfrak{R}$, $A_j$, and $\Delta C_j$ are defined as follows:
\begin{align}\label{eq:R2}
\mathfrak{R} = \frac{\Omega + \sum\nolimits_{j=1}^{J'}{A_j}- C_0\sum\nolimits_{j=1}^{J'}{N_j}}{\sum\nolimits_{j =1}^{J'}{\frac{N_j}{T_j}}},
\end{align}
\begin{align}\label{eq:Aj}
A_{j}\!=\!\left\{ \begin{array}{l}
 {N_{J'} C_{J'}},~~~~~~~~~~~~~~~~~~~~j=J';\\
 {N_j} C_j + \Delta C_j \sum\limits_{k=j+1}^{J'}{N_k},~~ j\le J'-1,\\
\end{array} \right.
\end{align}
\begin{align}\label{eq:deltaCj}
\Delta C_j = C_j -C_{j+1}.
\end{align}
\end{theorem}

\begin{IEEEproof}
Please refer to Appendix F.
\end{IEEEproof}

\emph{Remark.}
If $\mathbf{S}^*=\{S_{j}^*\}_{j\in \mathcal{J}'}$ is an non-decreasing sequence (i.e., C1 holds), {then} $\mathbf{S}^*$ is the solution of Problem~{2-1}. Nevertheless, the monotonicity constraint C1 may not hold in general {UAV's} type distributions. Based on \cite{5738226}, a dynamic VDD size assignment method is designed to cope with this issue through bunching and ironing.

\begin{theorem}\label{theorem2-7}
Define ${\tilde y}_n^*=\mathop {{\arg\max} }\nolimits_ { {y_n}} {\Gamma_n(y_n)}$ and $\Gamma_n(y)$ as a convex function of $y$, $\forall n=1,\cdots,N$. If $\tilde{y}_N^* \geq \tilde{y}_2^* \geq \cdots \geq \tilde{y}_1^*$ holds, we have $y_1^*=y_2^*=\cdots=y_N ^*$, where
\begin{align}\label{eq:4-1-theorem4}
\begin{gathered}
\{y_n^*\}=\arg\mathop{\max}\limits_{\{y_n\}} \sum\nolimits_{n = 1}^N {{\Gamma_n}({y_n})} , \forall n =1,\cdots,N \hfill \\
   \mathrm{s.t.}\; y_1 \geq y_2 \geq \cdots \geq y_N.
\end{gathered}
\end{align}
\end{theorem}

\begin{IEEEproof}
The detailed proof can refer to \cite{5738226}. As a single-variable optimization problem, the problem in (\ref{eq:4-1-theorem4}) can be efficiently solved by methods such as binary search.
\end{IEEEproof}

In Algorithm~\ref{Algorithm2-1}, a dynamically optimal VDD sequence allocation method with low-complexity is designed in lines 11-13 to iteratively search for sub-sequences that violate contractual feasibility and adjust them {to ensure contractual feasibility} by Theorem \ref{theorem2-7}.
Specifically, for any decreasing sub-sequence $\{S_{l}^*,S_{l+1}^*,\cdots,S_{m}^*\}\subseteq \mathbf{S}^*$, all its elements are dynamically adjusted by resolving the following single variable optimization problem:
\begin{align}\label{eq:OptSizeAssign}
S_{n}^* \!=\! \arg\mathop{\max}\limits_{S_{n}} &\sum\nolimits_{n = l}^m {{\frac{\varpi N_n}{T_n}\log \left(1 \!+\! S_n \right)} + \lambda_2 ({{A_n} S_{n} + N_n {C_0}})} \nonumber \\
&~~~- (\lambda_2 + 1) \Omega,\, \forall n =l,l\!+\!1,\!\cdots\!,m.
\end{align}
{Here, the definition of parameter $A_n$ can refer to Appendix~F.}

\emph{Remark.}
The above process in (\ref{eq:OptSizeAssign}) is repeated until all the sub-sequences in $\mathbf{S}^*$ obtained from (\ref{eq:OptimalSize1})--(\ref{eq:deltaCj}) are non-decreasing. After that, the optimal contracts $\Phi^*=\{T_{\max},\{S_j^*,R_j^*\}_{j\in \mathcal{J}}\}$ can be designed for all types of UAVs.

\begin{algorithm}[t!]\begin{small}
   \caption{\textbf{Budget-Constrained Optimal Contract Assignment in Partial Information Asymmetry}}\label{Algorithm2-1}
        \textbf{Input:} $\mathcal{J}'$, $N_j$, $\theta_j$, $\varpi$, $S_{\max}$, $C_0$, $T_{\max}$\;
        \textbf{Output:} Optimal contract ${\Phi}^*=\{T_{\max},\{{S}_j^{*},{R}_j^{*}\} _{j\in \mathcal{J}}\}$ \;
        \For{ $j \in \mathcal{J}\backslash \mathcal{J}'$}{
        Set ${S}_{j}^{*} = {R}_{j}^* = 0$\;
        }
        \For{ $j \in \mathcal{J}'$}{
            Calculate the relaxed optimal {contractual} VDD size strategy $\widetilde{{S}}_{j}^{*}$ via Theorem 4\;
            \If{ $\widetilde{{S}}_{j}^{*} > S_{\max}$}{
                Set $\widetilde{{S}}_{j}^{*} = S_{\max}$\;
            \ElseIf{ $\widetilde{{S}}_{j}^{*} < 0$}{
                Set $\widetilde{{S}}_{j}^{*} = 0$\;
                }
            }
        }
        \While{VDD sequence$\{\widetilde{{S}}_{j}^{*}\}_{j\in \mathcal{J}'}$ does not satisfy the {contractual} feasibility}{
            Search for one of the {sub-sequences} $\{\widetilde{{S}}_{l}^{*},\widetilde{{S}}_{l+1}^{* },\cdots,\widetilde{{S}}_{m}^{*}\} \subseteq \{\widetilde{{S}}_{j}^{*}\}_{j\in \mathcal{J}'}$\;
            Dynamically adjust the infeasible {sub-sequence} by (\ref{eq:OptSizeAssign})\;
        }
        \For{$j \in \mathcal{J}'$}{
            Compute the optimal {contractual} reward strategy $R_{j}^{*}=R_{j}^{*}\left( \mathbf{S}^* \right )$ by (\ref{eq:4-1-OptimalPrice})\;
        }
\end{small}
\end{algorithm}

Algorithm \ref{Algorithm2-1} describes the optimal contract design process in the UAV honeypot game under partial information asymmetry and budget limits. First, in lines 3--4, the GCS sets up a zero-payment contract for non-participating UAVs and UAVs that cannot transmit VDD in time. Next, in lines 5--10, for UAVs involved in honeypot defense, the GCS calculates the optimal contract VDD size strategy $\widetilde{{S}}_{j}^{*}$ according to (\ref{eq:OptimalSize1})--(\ref{eq:deltaCj}). Lines 11--13 represent the dynamic allocation process of the optimal VDD size sequence.
After obtaining the optimal VDD size sequence ${\mathbf{S}}^*$, in lines 14--15, the GCS calculates the optimal contract reward $R_{j}^{*}$ by (\ref{eq:4-1-OptimalPrice}).
In each round of collaborative defense based on honeypot game, each participating UAV uploads its VDD data according to the contract data size and receives the corresponding contract reward from the GCS after completing data transmission in time.

{\emph{Complexity and Convergence Analysis.}
In Algorithm~\ref{Algorithm2-1}, the parts to be iterated only exist in the while loop (i.e., lines 11-13).
For the while loop, the maximum number of iterations is $J{'}-1$. In other words, Algorithm~\ref{Algorithm2-1} is guaranteed to be converged within $J{'}-1$ iterations. In each iteration of the while loop, the search of an infeasible sub-sequence incurs a maximum computation overhead of $\mathcal{O}(\log J{'})$, while the adjustment of an infeasible sub-sequence yields $\mathcal{O}(J{'})$ as the intermediate parameters in solving (\ref{eq:OptSizeAssign}) can be pre-computed. For the rest of Algorithm~\ref{Algorithm2-1} except lines 11-13, as $\mathfrak{R}$, $A_j$, and $\Delta C_j$ in (\ref{eq:R2})--(\ref{eq:deltaCj}) can be pre-computed, it also yields a $\mathcal{O}(J{'})$ overhead for optimal contract calculation.
As such, the overall computational complexity of Algorithm~\ref{Algorithm2-1} is $\mathcal{O}(J{'}^{2})$.}

\begin{theorem}\label{theorem2-9}
The optimal contracts derived in Algorithm~\ref{Algorithm2-1} satisfy contractual fairness.
\end{theorem}

\begin{IEEEproof}
According to Theorem \ref{theorem2-5}, any UAV that does not participate in honeypot data sharing will receive a non-positive payoff. Since the optimal contracts satisfy IR constraints, the payoff of an honest UAV is always non-negative and no less than the case when it does not participate. Hence, the designed optimal contract satisfies participation fairness.
According to Theorem \ref{theorem2-5}, for every type-$j$ UAV, its optimal contract reward $R_{j}^*({\mathbf{S}}^*)$ increases with the increase of the contract VDD size $S_{j}^{*}$. Furthermore, for non-participating UAVs, the proposed contract mechanism enforces a zero-payment strategy. Hence, the optimal contracts guarantee reward fairness. According to Definition~8, the obtained optimal contracts in Algorithm~\ref{Algorithm2-1} satisfy contractual fairness.
\end{IEEEproof}

\section{Optimal Dynamic Contract Design in Complete Information Asymmetry}\label{sec:DynamicSolution}
In this section, we design the optimal dynamic contract in complete information asymmetry. Different from the partial information asymmetry scenario, the GCS has no prior knowledge of the UAVs' private types under the complete information asymmetry. {We first formulate a Markov game under complete information asymmetry in Sect.~\ref{MarkovGame}. Then, in Sect.~\ref{PHCRewardGCS} and Sect.~\ref{PHCVDDUAV},} both the GCS and UAVs apply {the} policy hill-climbing (PHC) learning, a model-free RL technique, to make optimal reward and VDD size strategies in the dynamic contract through trials, without explicitly knowing UAVs' private parameters (e.g., {UAVs'} type distribution).

{\subsection{Multi-Agent Markov Game Formulation under Complete Information Asymmetry}\label{MarkovGame}
Under the complete information asymmetry, both the GCS and UAVs can make repeated interactions and exploit historical interacting experience to derive the optimal strategies. To facilitate the analysis, we assume that the evolution of GCS/UAV's state in the future only depends upon the present state instead of the past ones \cite{9809923}. As such, the strategy-making processes of the GCS and UAVs can be modelled as finite MDPs. Then, the one-shot honeypot game $\mathcal{G}$ between the GCS and UAVs can be extended as a Markov game\footnotemark[3] with repeated interactions under the complete information asymmetry scenario, which is defined as below.\footnotetext[3]{{In game theory, the Markov game (or called stochastic game) is introduced by Lloyd Shapley, which consists of a sequence of non-cooperative stage games that are repeated over time for one or more players \cite{9809923}. Essentially, Markov games generalize Markov decision processes (MDPs) to multiple interacting players (i.e., decision makers) \cite{6457437}.}}}
\begin{definition}[Multi-Agent Markov Game]
Under repeated interactions between UAVs and the GCS, the one-shot honeypot game $\mathcal{G}$ can be extended as a multi-agent Markov game $\mathcal{G}'=\left\{T_e,\{G,\mathcal{J}\},\{W_j^t,\tilde{W}_j^t\},\{{\bf{P}}_G, {\tilde{\bf{P}}}_j\},\{R_j,S_j\},\{\mathcal{U}_G,\{\mathcal{U}_j\}\}\right\}$, which includes the following main components:
\begin{itemize}
  \item \emph{Players.} In game $\mathcal{G}'$, (i) UAVs with diverse private types in the set $\mathcal{J}$ and (ii) the GCS $G$ are the players.
  \item \emph{States.} (i) For the GCS, its observed system state vector at time slot $t$ is denoted as ${\bf{W}}^t = \left(W_1^t,\cdots,W_{J'}^t\right)$, which consists of the previous VDD size of each type of UAV, i.e., ${\bf{W}}^t = {\bf{S}}^{t-1}$. (ii) For each type-$j$ UAV, its observed state at time slot $t$ is denoted as $\tilde{W}_j^t$, which contains the GCS's previous reward, i.e., $\tilde{W}_j^t = {R}_{j}^{t-1}$.
  \item \emph{Actions.} (i) The GCS uniformly quantizes the reward action into $A+1$ levels, i.e., ${R}_j \in \mathcal{A}=\{\frac{a}{A}\cdot R_{\max}\}_{0 \le a \le A}$, where $R_{\max}$ is the maximum contractual reward that the GCS pays to a UAV. (ii) Each type-$j$ UAV uniformly quantizes its contractual VDD size strategy into $B+1$ levels, i.e., $S_j \in \mathcal{B}=\{\frac{b}{B}\cdot S_{\max} \}_{0 \le b \le B}$, where $S_{\max}$ is the maximum contractual VDD size for a UAV.
  \item \emph{State Transitions.} (i) For the GCS, its state transition vector is denoted as ${\bf{P}}_G = \{{\bf{P}}_j\}_{j \in \mathcal{J}}$. The state transition matrix is ${\bf{P}}_j = [p_{l,k}^{R_j}]^{(B+1)\times(B+1)}$, where
      \begin{align}
      p_{l,k}^{R_j} = \Pr(S_l|S_k,R_j), \forall {0 \le l,k \le B},
      \end{align}
      and $\sum_{l=0}^{B}{p_{l,k}^{R_j}} = 1, \forall S_k \in \mathcal{B}$.\\
      (ii) For each type-$j$ UAV, its state transition is denoted as ${\tilde{\bf{P}}}_j = [{\tilde{p}}_{l,k}^{S_j}]^{(A+1)\times(A+1)}$, where
      \begin{align}
      {\tilde{p}}_{l,k}^{S_j} = \Pr(R_l|R_k,S_j), \forall {0 \le l,k \le A},
      \end{align}
      and $\sum_{l=0}^{A}{\tilde{p}}_{l,k}^{S_j} = 1, \forall R_k \in \mathcal{A}$.
  \item \emph{Rewards.} The immediate rewards to type-$j$ UAV and the GCS are their stage payoffs (or utilities), which are denoted as ${\mathcal{U}_j}\left(\tilde{W}_j^t, S_j^t\right)$ and ${\mathcal{U}_G}\left({\bf{W}}^t, {\bf{R}}^t\right)$, respectively.
\end{itemize}
Note that the Markov game $\mathcal{G}'$ is repeated, i.e., it consists of multiple stage games that are repeated over time. Here, $T_e$ is the maximum interaction times. The Markov transition probabilities are determined once the policies of all players are optimized \cite{6457437}.
\end{definition}

\subsection{PHC-based Reward Strategy of The GCS}\label{PHCRewardGCS}
Let $\mathcal{{Q}}\left({\bf{W}}^t, {\bf{R}}^t\right)$ represent the GCS's Q-function (i.e., expected long-term discounted {sum of the stage utilities}) of each state-action pair, which is updated based on the iterative Bellman equation:
\begin{align}\label{eq:6-1}
\mathcal{{Q}}\left({W}_j^t, R_j^t\right) &\leftarrow (1-\kappa_1)\mathcal{{Q}}\left({W}_j^t, R_j^t\right) + \kappa_1 \left\{ {\mathcal{U}_G}\left({W}_j^t, R_j^t\right) \right. \nonumber \\
&~\left. {+ \varphi_1 \mathop {\max }\limits_{R_j} \mathcal{{Q}}\left(W_j^{t+1}, R_j^{t+1}\right)} \right\}, \forall j \in \mathcal{J}',
\end{align}
where $\kappa_1$ is the learning rate, and $\varphi_1$ is the discount factor. $W_j^{t+1}$ is the new state of type-$j$ UAV at time slot $t+1$, which is transferred from state $W_j^{t}$ with action $R_j^t$.

To tradeoff the exploration and exploitation in PHC, the mixed-strategy table $\pi\left({\bf{W}}^t, {\bf{R}}^t\right)$, {i.e., the policy of the GCS,} is updated by increasing the chance of behaving greedily by a small value $\rho_1$, and lowering other chances by $-\frac{\rho_1}{A+1}$, i.e.,
\begin{align}\label{eq:6-3}
\pi &\left(W_j^t, R_j^t\right) \gets \pi \left(W_j^t, R_j^t\right)\nonumber \\
&+\left\{ \begin{array}{cl}
	\rho _1,&\mathrm{if}\ R_j^{t}=\arg\max_{R_j}\mathcal{Q}\left( W_j^{t},R_j \right);\\%[0.02cm]
	-\frac{\rho _1}{A+1},&\mathrm{otherwise}.
\end{array} \right.
\end{align}
The GCS opts its contractual reward strategy $R_j^t,\forall j\in \mathcal{J}'$ based on the above mixed-strategy table, i.e.,
\begin{align}\label{eq:6-4}
\Pr\big( R_j^{t} = \hat{R}_j \big) = \pi \big( W_j^t,\hat{R}_j \big) ,\forall \hat{R}_j\in \mathcal{A}.
\end{align}
The hotbooting PHC-based optimal contractual reward strategy-making process of the GCS is summarized in lines 4--10 in Algorithm~\ref{Algorithm2}.

\begin{algorithm}[t!]\begin{small}
   \caption{\textbf{Optimal Dynamic Contract with Hotbooting PHC in Complete Information Asymmetry}}\label{Algorithm2}
        \textbf{Initialize: }$\kappa_1$, $\kappa_2$, $\varphi_1$, $\varphi_2$, $\rho_1$, $\rho_2$, ${\bf{W}}^0$, $\tilde{W}_j^0$, $A$, $B$\;
        Perform hotbooting process and obtain ${\mathcal{{Q}}} = {\mathcal{{Q}}}_p$, ${\pi} = {\pi}_p$, $\tilde{{\mathcal{{Q}}}} = \tilde{{\mathcal{{Q}}}}_p$, $\tilde{{\pi}} = \tilde{{\pi}}_p$\;
        \For{$t = 1, 2,\cdots,T_e $}{
            \textbf{Layer 1: Hotbooting PHC-Based Reward Strategy of The GCS}\;
            Set system state vector ${\bf{W}}^t = {\bf{S}}^{t-1}$\;
            Select payment action vector ${\bf{R}}^{t} {= \left(R_1^t,\cdots,R_{J'}^t\right)}$ by (\ref{eq:6-4})\;
            Observe and evaluate {UAVs'} VDD size vector ${\bf{S}}^{t}{= \left(S_1^t,\cdots,S_{J'}^t\right)}$\;
            Evaluate {the} reward ${\mathcal{U}_G}\left({W}_j^t, R_j^t\right)$ by (\ref{eq:utility-GCS})\;
            Update $\mathcal{{Q}}\left({W}_j^t, R_j^t\right)$ by (\ref{eq:6-1})\;
            Update $\pi \left(W_j^t, R_j^t\right)$ by (\ref{eq:6-3})\;
            \textbf{Layer 2: Hotbooting PHC-Based VDD Size Strategy of Each Type of UAV}\;
            Set system state $\tilde{W}_j^t = {R}_{j}^{t-1}$\;
            Select VDD size action $S_j^t$ by (\ref{eq:6-8})\;
            Observe the {GCS's} payment ${R}_j^{t}$\;
            Evaluate {the} reward ${\mathcal{U}_j}\left(\tilde{W}_j^t, S_j^t\right)$ by (\ref{eq:utility-UAV})\;
            Update $\tilde{\mathcal{{Q}}}\left(\tilde{W}_j^t, S_j^t\right)$ by (\ref{eq:6-5})\;
            Update $\tilde{\pi}\left(\tilde{W}_{j}^t, S_{j}^t\right)$ by (\ref{eq:6-7})\;
        }
    \end{small}
\end{algorithm}

\subsection{PHC-based VDD Size Strategy of The UAV}\label{PHCVDDUAV}
Let $\tilde{\mathcal{{Q}}}\left(\tilde{W}_j^t, S_j^t\right)$ denote the Q-function of type-$j$ UAV. Similarly, the Q-function is updated by the iterative Bellman equation:
\begin{align}\label{eq:6-5}
\tilde{\mathcal{{Q}}}\left(\tilde{W}_j^t, S_j^t\right) &\leftarrow (1-\kappa_2)\tilde{\mathcal{{Q}}}\left(\tilde{W}_j^t, S_j^t\right) + \kappa_2 \left\{ {\mathcal{U}_j}\left(\tilde{W}_j^t, S_j^t\right) \right. \nonumber \\
&\left. {+ \varphi_1 \mathop {\max }\limits_{S_j} \tilde{\mathcal{{Q}}} \left( \tilde{W}_{j}^{t+1}, S_{j}^{t+1} \right) } \right\},\forall j \in \mathcal{J}',
\end{align}
where $\kappa_2$ is the learning rate, and $\varphi_2$ is the discount factor.

Similarly, the {policy of type-$j$ UAV defined by its} mixed-strategy table $\tilde{\pi}\left(\tilde{W}_{j}^t, S_{j}^t\right)$ in PHC is updated by
\begin{align}\label{eq:6-7}
\tilde{\pi}&\left(\tilde{W}_{j}^t, S_{j}^t\right) \gets \tilde{\pi}\left(\tilde{W}_{j}^t, S_{j}^t\right)\nonumber \\
&+\left\{ \begin{array}{cl}
	\rho _2,&\mathrm{if}\ S_j^{t}=\arg\max_{S_j}\tilde{\mathcal{{Q}}}\left(\tilde{W}_{j}^t, S_j\right);\\%[0.02cm]
	-\frac{\rho _2}{B+1},&\mathrm{otherwise}.
\end{array} \right.
\end{align}
Here, $\rho _2$ is a small positive value. According to the mixed-strategy table, each type-$j$ UAV ($j\in \mathcal{J}'$) chooses its VDD size strategy $S_{j}^t$ with the following chance:
\begin{align}\label{eq:6-8}
\Pr\left( S_j^{t} = \hat{S}_{j} \right) = \pi \left(\tilde{W}_{j}^t,\hat{S}_{j}\right) ,\forall \hat{S}_{j}\in \mathcal{B}.
\end{align}
The hotbooting PHC-based optimal contractual VDD size strategy-making process of each type of UAV is summarized in lines 11--17 in Algorithm~\ref{Algorithm2}.

\emph{Remark.}
The above two-layer strategy-making process is repeated between the GCS and each type of UAV in $\mathcal{J}'$ until the strategies of both sides converge to stable values.

%\vspace{-2mm}
\subsection{Hotbooting PHC for Practical Deployment}
To speed up the convergence rate, a hotbooting technique is employed for both sides by learning from similar scenarios in an offline manner for efficient initialization of the Q-table and mixed-strategy table. Specifically, as shown in line 2 in Algorithm~\ref{Algorithm2}, by exploiting $p$ numbers of historical interactions conducted in similar scenarios before the game starts, the hotbooting process outputs ${\mathcal{{Q}}}_p$ and $\tilde{{\mathcal{{Q}}}}_p$ as the initial Q-tables, and outputs ${\pi}_p$ and $\tilde{{\pi}}_p$ as the initial mixed-strategy tables. Thereby, the inefficient random explorations in traditional PHC learning with all-zero initialization of Q-value and mixed-strategy table can be mitigated. The overall computational complexity of Algorithm~\ref{Algorithm2} yields $\mathcal{O}(J' \times T_e)$, where $T_e$ means the maximum interaction times and $J'$ is the number of types of participating UAVs.

\begin{figure}[!t]\setlength{\abovecaptionskip}{-0.1cm}\vspace{-1.5mm}
\centering
  \includegraphics[width=7.95cm]{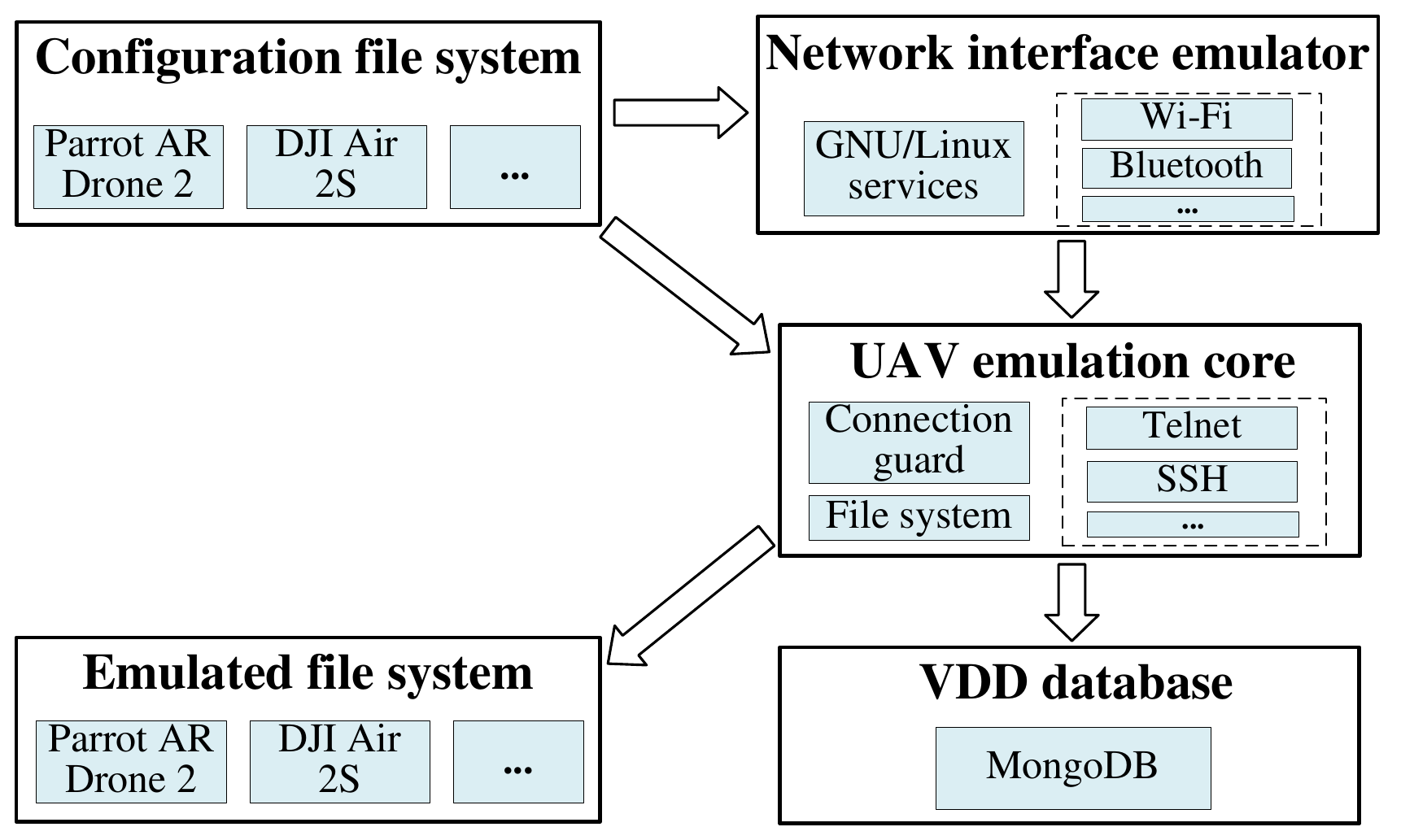}
  \caption{{Implementation architecture of the UAV honeypot prototype.}}\label{fig:UAVhoneypot-implement}\vspace{-3.mm}
\end{figure}

\section{PERFORMANCE EVALUATION}\label{sec:SIMULATION}
%This section first introduces the simulation setup, followed by the numerical results and discussions.

\subsection{Simulation Setup}\label{subsec:evalution1}
We consider a simulation area of $200\times200\times80\, \mathrm{m}^3$ with one GCS and $I=10$ uniformly placed Parrot AR Drone 2.0 UAVs. {The height of UAV is fixed and is randomly located between $[20,80]$m, and UAV's maximum velocity is set as $20$ m/s. UAV's trajectory is a circle whose center is the GCS and the radius is the initial horizontal distance between the UAV and GCS.} Each UAV is embedded with a honeypot system and communicates with the GCS and other UAVs via Wi-Fi communications. {The UAV honeypot is implemented on a Raspberry Pi 2 Model B with Raspbian operating system (OS), 1GB RAM, and 900MHz ARM Cortex-A7 quad-core CPU, and its software core is implemented with the Twisted framework using Python 2.7.}
The Telnet attack \cite{TelnetParrotDrone} is considered in the simulation, {where the adversary can have root privilege on the file system after successfully establish a Telnet connection with the UAV/honeypot.}

Similar to the HoneyDrone project \cite{Daub2018HoneyDrone}, {the UAV honeypot shares the same IP address with the real UAV and it mainly consists of five components: configuration file system (CFS), network interface emulator (NIE), UAV emulation core (UEC), emulated file system (EFS), and VDD database, as shown in Fig.~\ref{fig:UAVhoneypot-implement}.
Here, CFS specifies the network interface in NIE and the file system in EFS after activating the honeypot. NIE is to set up the network interface (e.g., Wi-Fi) in the honeypot. UEC handles incoming connections via the connection guard (CG), emulate specific protocols (e.g., Telnet) in NIE, and continuously monitors the data traffic.}
Specifically, the honeypot reads UAV profiles from the CFS to produce the EFS, emulates UAV's radio interfaces in NIE, and offers low to medium interactions with adversaries for Wi-Fi protocols via UEC. The honeypot's captured VDD (including {attackers'} IP addresses, port numbers, connection types, commands, and timestamps) is recorded into a local MongoDB database.
The medium interaction is set as the default option of the UAV honeypot prototype.

The GCS requests VDD from UAVs every $6$ seconds, with a maximum communication delay of $2$ seconds and a default system budget $\Omega=460$.
For simplicity, UAVs' types are assumed to be uniformly distributed. The lower and upper bounds of UAV's marginal VDD cost are set as $0.01$ and $1$, respectively.
The A2A/A2G channel parameters are set based on works \cite{8624565,9115898}.
Specifically, we set $\iota = 2$, $\iota_1=12$, $\iota_2=0.135$, $\kappa_{\mathrm{LoS}} =1$, $\kappa_{\mathrm{NLoS}} =20$, $ P_{i}^{\text{Tr}} =23$ dBm, ${\varphi^2} =-96$ dBm, ${B^{\mathrm{A2G}}}=1$ MHz, ${B^{\mathrm{A2A}}} =0.25$ MHz.
For the utility model, we set $\varpi=6$, $C_0 = 1$, $T_{\max} = 2$ seconds, $S_{\max} = 300$ bytes.
For the PHC learning, we set $\kappa_1=\kappa_2=0.7$, $\varphi_1=\varphi_2=0.8$, $\rho_1=\rho_2=0.01$.
Simulation parameters are summarized in Table~\ref{table2-simu}.

\begin{table}[!tp]\begin{center}\setlength{\abovecaptionskip}{-0.0cm}
        \caption{Simulation Parameters}\label{table2-simu}
        \begin{tabular}{cc|cc}
        \hline
        \textbf{{Param}} & \textbf{{Value}} & \textbf{{Param}} & \textbf{{Value}} \\ \hline
            $I$ &$10$ &$J'$ &$10$  \\
            $z_i$ &$[20,80]$ m &$V_{\max}^i$ &$20$ m/s \\
            $S_{\max}$ &$300$ bytes &$T_{\max}$ &$2$ seconds\\
            $C_0 $ &$1$ & $C_{j}$ &$[0.01,1]$ \\
%            $\overline{R}$ &$480$ &$D_G$ &$800$ bytes \\
            $D_G$ &$800$ bytes &$\varpi$ &$6$ \\
            $\iota$ &$2$ &$ \delta_{th} $&$10$ dB\\
            $ \kappa_{\mathrm{LoS}} $&$1$ &$ \kappa_{\mathrm{NLoS}} $&$20$  \\
            $ \iota_1$ & $12$ &$\iota_2$ &$0.135$   \\
            $ P_{i}^{\text{Tr}} $&$23$ dBm &$ {\varphi^2} $&$-96$ dBm  \\[0.03cm]
            $ {B^{\mathrm{A2G}}}$&$1$ MHz&$ {B^{\mathrm{A2A}}} $&$0.25$ MHz \\
            $ \kappa_1$&$0.7$&$ \kappa_2 $&$0.7$ \\
            $ \varphi_1$&$0.8$&$ \varphi_2 $&$0.8$ \\
            $ \rho_1$&$0.01$&$ \rho_2 $&$0.01$ \\
            \hline
        \end{tabular}\end{center}
\end{table}

The following three conventional contract approaches are used for performance comparison with the proposed scheme.
\begin{itemize}
  \item \emph{Complete information contract}. In this ideal scenario, the GCS knows the private type of each UAV, and only IR constraints should be met in optimal contract design via Eqs. (\ref{eq:CIoptS}) and (\ref{eq:CIoptR}).
  \item \emph{Linear contract}. The reward offered by the GCS is in direct proportion to UAV's shared VDD size in this contract, i.e., $R_j=\mu_{G}\times S_j$, $\forall j \in \mathcal{J}'$, where $\mu_G$ is the unit payment per VDD size. Here, we set $\mu_G=\max\{C_j,\forall j \in \mathcal{J}'\}$, i.e., $\mu_G=C_1$.
  \item \emph{Uniform contract}. In this contract, the GCS applies a single uniform contract item for all types of UAVs, i.e., $\Phi_j=\{S_1^*,R_1^*\}$, $\forall j \in \mathcal{J}'$.
\end{itemize}

\begin{table}
\setlength{\abovecaptionskip}{-0.0cm}
{
\caption{Evaluation on CPU Utilization of The UAV Honeypot}
\label{simu:CPUtable}\centering~~~~~~~
\begin{tabular}{|c|c|}
\hline
\textbf{Status}                  & \textbf{CPU utilization ratio} \\ \hline
Idle                             & Max. 49.4\%                      \\ \hline
Running UAV honeypot services    & Ave. 15.2\% above idle           \\ \hline
1 Telnet connection to attacker  & Ave. 1.6\% above idle          \\ \hline
4 Telnet connections in parallel & Ave. 6.6\% above idle          \\ \hline
\end{tabular}}
\end{table}

\begin{table}
\setlength{\abovecaptionskip}{-0.01cm}
\caption{Comparison of Defense Effectiveness, Compromised Rate of UAVs, and Resource Consumption on Different Interaction Levels of UAV Honeypots}
\label{simu:Defensetable}\centering
\resizebox{1.01\linewidth}{!}{
\begin{tabular}{|c|c|c|c|}
\hline
\textbf{Interaction Levels} &
  \textbf{\begin{tabular}[c]{@{}c@{}}Attack Det. \\ Rate\end{tabular}} &
  \textbf{\begin{tabular}[c]{@{}c@{}}Compromised \\ Rate of UAVs\end{tabular}} &
  \textbf{\begin{tabular}[c]{@{}c@{}}Resource Consumption\\ for 1 Telnet connection\end{tabular}} \\ \hline
\textbf{\begin{tabular}[c]{@{}c@{}}Cooperative Medium-\\interaction Honeypots\end{tabular}} &
  91.8\% &
  0\% &
  \begin{tabular}[c]{@{}c@{}}Ave. 1.6\% above;\\ Run on Raspberry Pi 2\end{tabular} \\ \hline
\textbf{\begin{tabular}[c]{@{}c@{}}Cooperative High-\\interaction Honeypots\end{tabular}} &
  93.6\% &
  3.7\% &
  Run on Physical Server \\ \hline
\end{tabular}}
\end{table}

\subsection{Numerical Results}\label{subsec:evalution2}
{We first evaluate the CPU utilization of our UAV honeypot under different operations in Table~\ref{simu:CPUtable}. We start with the CPU utilization measurement in an idle Raspbian OS in Raspberry, then incrementally activate UAV honeypot services, and lastly connect simulated adversaries via Telnet and interact with them inside the honeypot. All these simulations are conducted for 40 times. As seen in Table~\ref{simu:CPUtable}, the CPU utilization of the idle Raspbian OS reaches its maximum value of 49.4\% before running honeypot services. After activating the honeypot services, it brings about an average 15.2\% of CPU utilization above the idle status. In this process, the NIE establishment and Telent protocol emulation constitute the most of the CPU utilization. In addition, for every adversarial Telnet connection to the UAV honeypot, the CPU utilization increases an additional 1.6\% on average. It can be concluded that the UAV honeypot is able to support multiple parallel connections/interactions with adversaries, without incurring significant overheads and performance degradation to the battery-powered UAV systems.}

Then, we evaluate attack detection rate, compromised rate of UAVs, and resource consumption on different interaction levels of UAV honeypots in Table~\ref{simu:Defensetable}. %In both honeypot configurations, UAV honeypots work collaboratively, and their captured VDD is shared among all the UAVs.
This experiment is repeated 50 times.
Here, the high-interaction honeypot simulates a real UAV system including its OS and software, and it can provide more detailed attack information. However, due to its resource-intensive nature, it is deployed on a physical server, rather than being integrated with the flying UAV. As shown in Table~\ref{simu:Defensetable}, the high-interaction honeypot achieves the highest attack detection rate (i.e., 93.6\%), but it also results in the highest compromised rate of UAVs (i.e., 3.7\%) and the highest resource consumption.
It is because the higher interaction honeypot provides more in-depth attack information to facilitate attack defense, but also increases the risk of being compromised by Telnet attackers. Furthermore, as the honeypot serves as an additional workload and only offer limited interactions in low/medium-interaction settings, it is difficult to invade the real UAV even if the UAV honeypot is compromised.
In summary, medium-interaction UAV honeypots offer a desirable attack detection rate (which is near to the high-interaction one), zero compromised rate of UAVs, and low CPU utilization rate on Raspberry Pi 2. Additionally, for low/medium-interaction UAV honeypots, the defensive effectiveness can be further enhanced by deploying high-interaction honeypots on the GCS and obtaining latest VDD from external security service providers.

\begin{figure*}[!tbp]\setlength{\abovecaptionskip}{-0.05cm}\vspace{-0.2cm}
\begin{minipage}[t]{0.32\textwidth}
\centering
    \includegraphics[height=4.8cm,width=\textwidth]{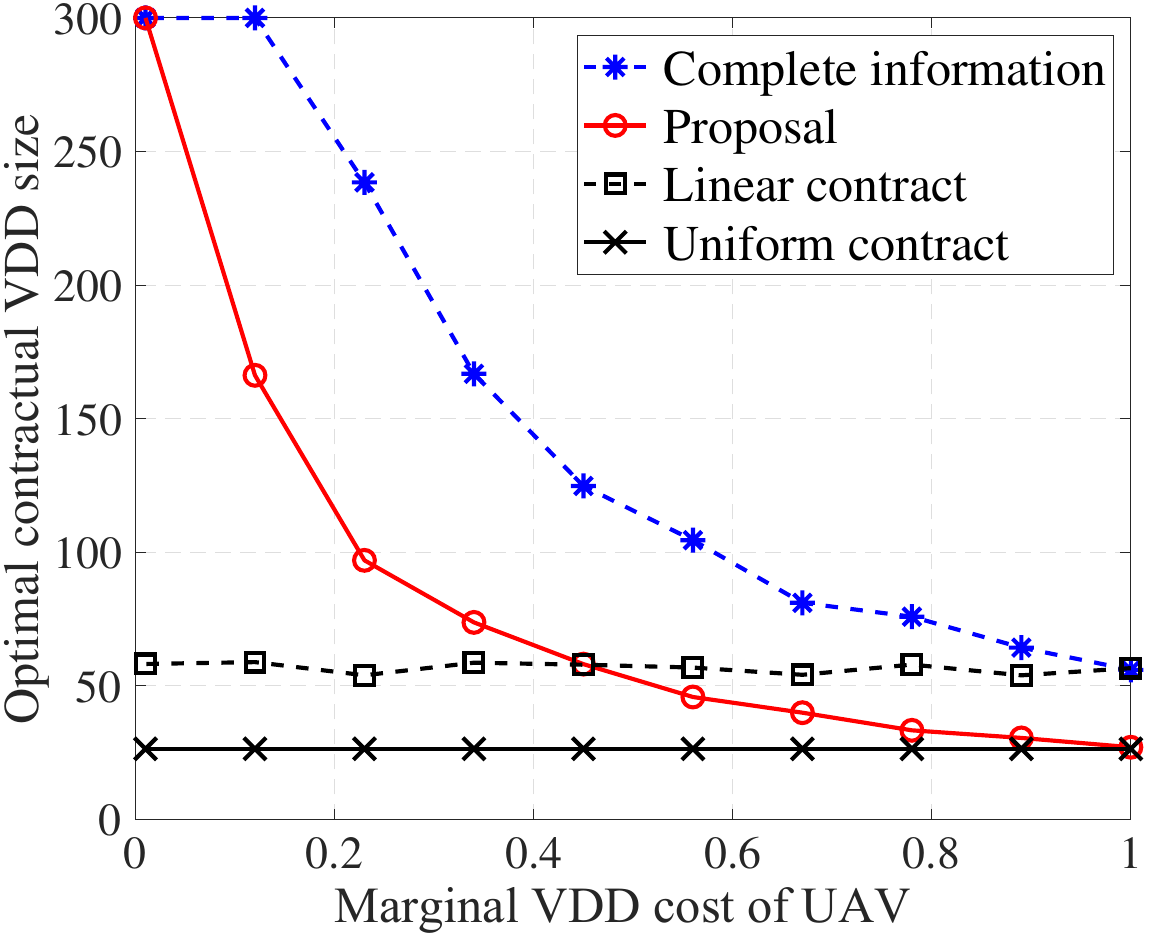}
    \caption{Optimal contractual VDD size vs. marginal VDD cost of UAV in the proposed scheme under partial information asymmetry, compared with other three contract approaches.}\label{simufig1}
\end{minipage}~~~~
\begin{minipage}[t]{0.32\textwidth}
\centering
    \includegraphics[height=4.8cm,width=\textwidth]{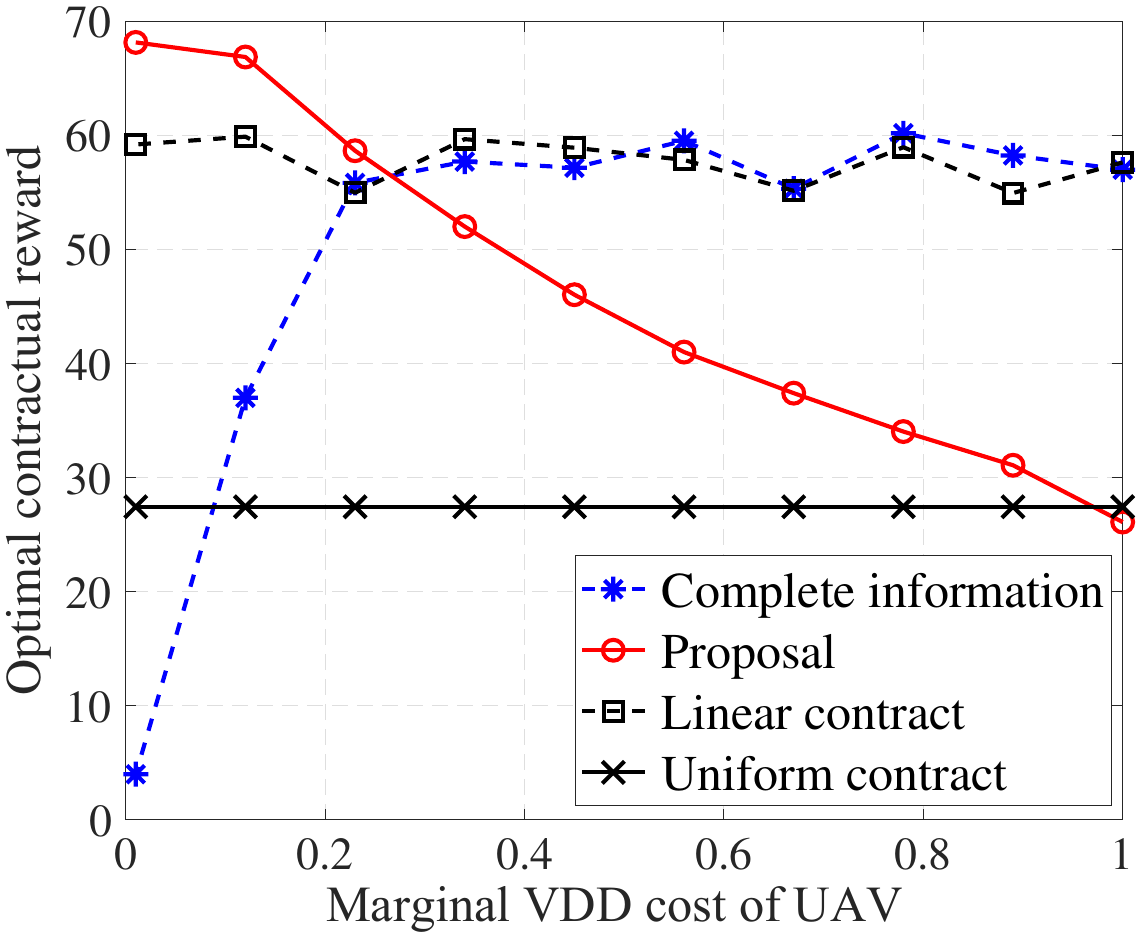}
    \caption{Optimal contractual reward vs. marginal VDD cost of UAV in the proposed scheme under partial information asymmetry, compared with other three contract approaches.}\label{simufig2}
\end{minipage}~~~~
\begin{minipage}[t]{0.32\textwidth}
\centering
    \includegraphics[height=4.8cm,width=\textwidth]{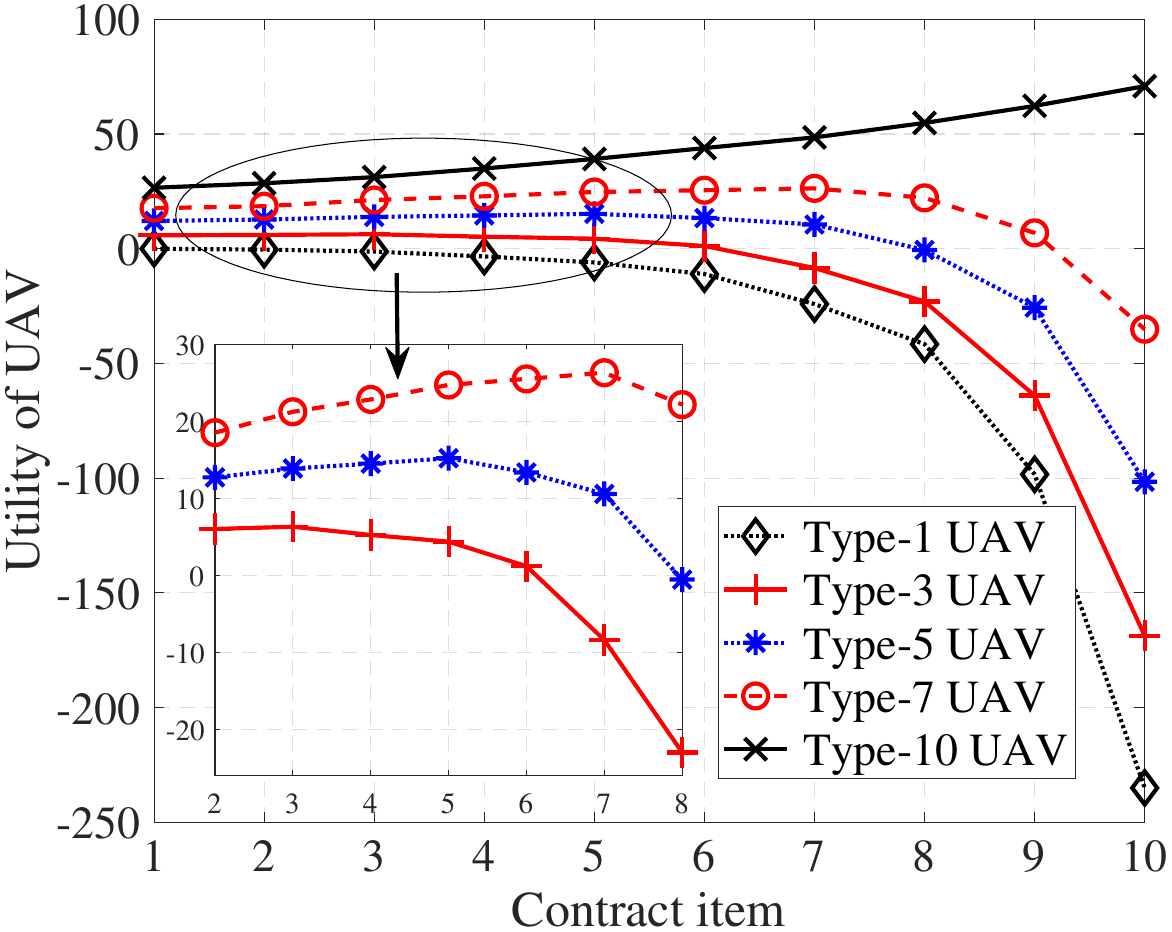}
    \caption{The utilities of different types of {UAVs} when selecting different contract items in the optimal contract under partial information asymmetry.}\label{simufig3}
\end{minipage}%\vspace{-0.2cm}
\end{figure*}

\begin{figure*}[!tbp]\setlength{\abovecaptionskip}{-0.05cm}\vspace{-0.2cm}
\begin{minipage}[t]{0.32\textwidth}
\centering
    \includegraphics[height=4.8cm,width=\textwidth]{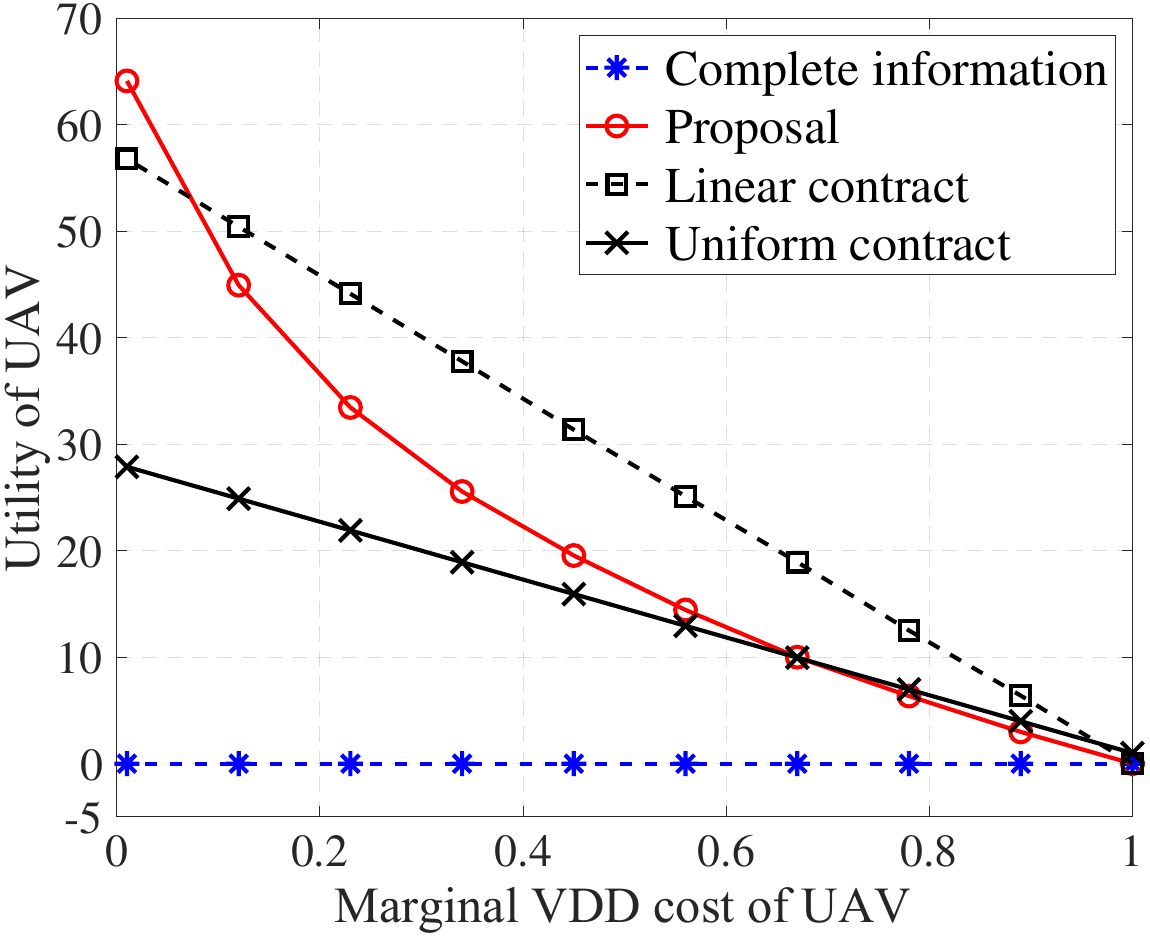}
    \caption{Utility of UAV vs. marginal VDD cost of UAV in the proposed scheme under partial information asymmetry, compared with other three contracts.}\label{simufig4}
\end{minipage}~~~~
\begin{minipage}[t]{0.32\textwidth}
\centering
    \includegraphics[height=4.8cm,width=\textwidth]{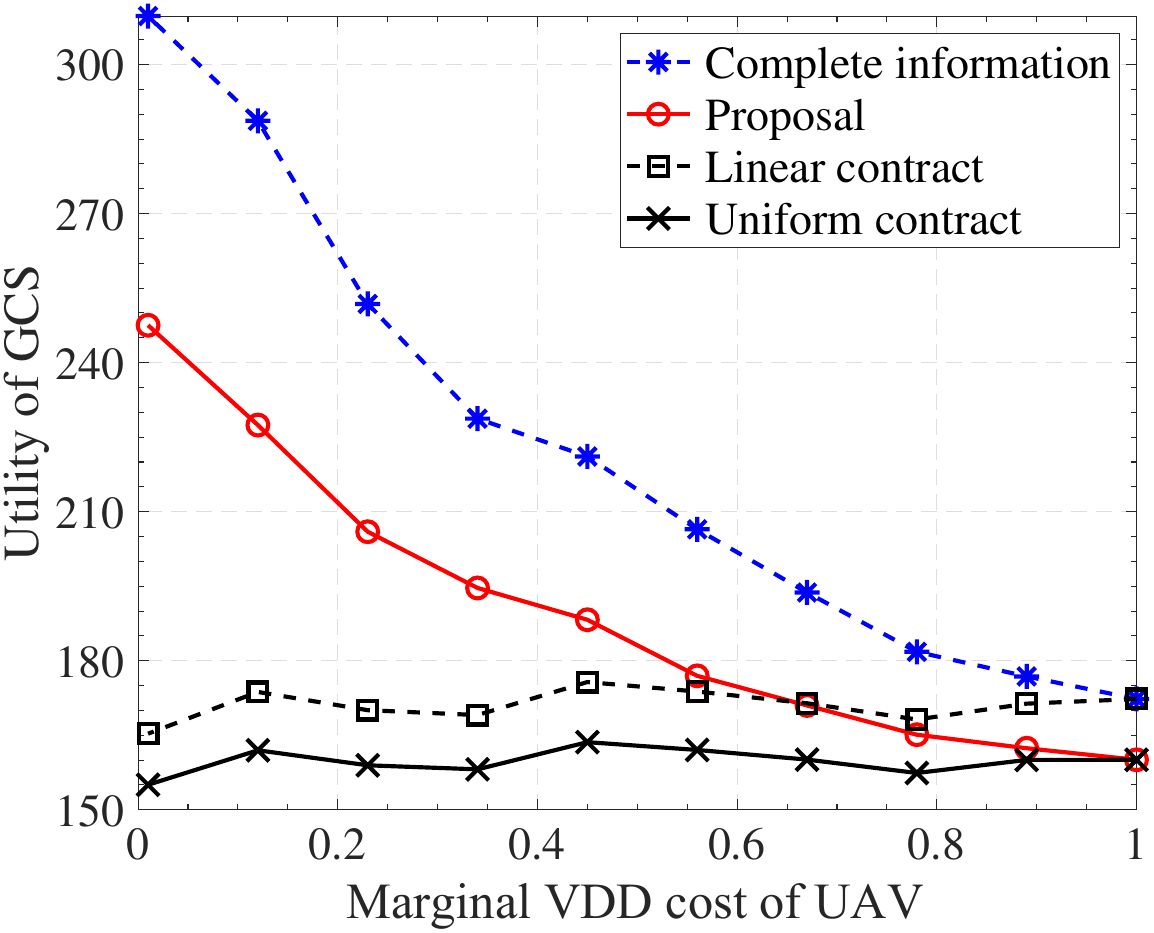}
    \caption{Utility of GCS vs. marginal VDD cost of UAV in the proposed scheme under partial information asymmetry, compared with other three contracts.}\label{simufig5}
\end{minipage}~~~~
\begin{minipage}[t]{0.32\textwidth}
\centering
    \includegraphics[height=4.8cm,width=\textwidth]{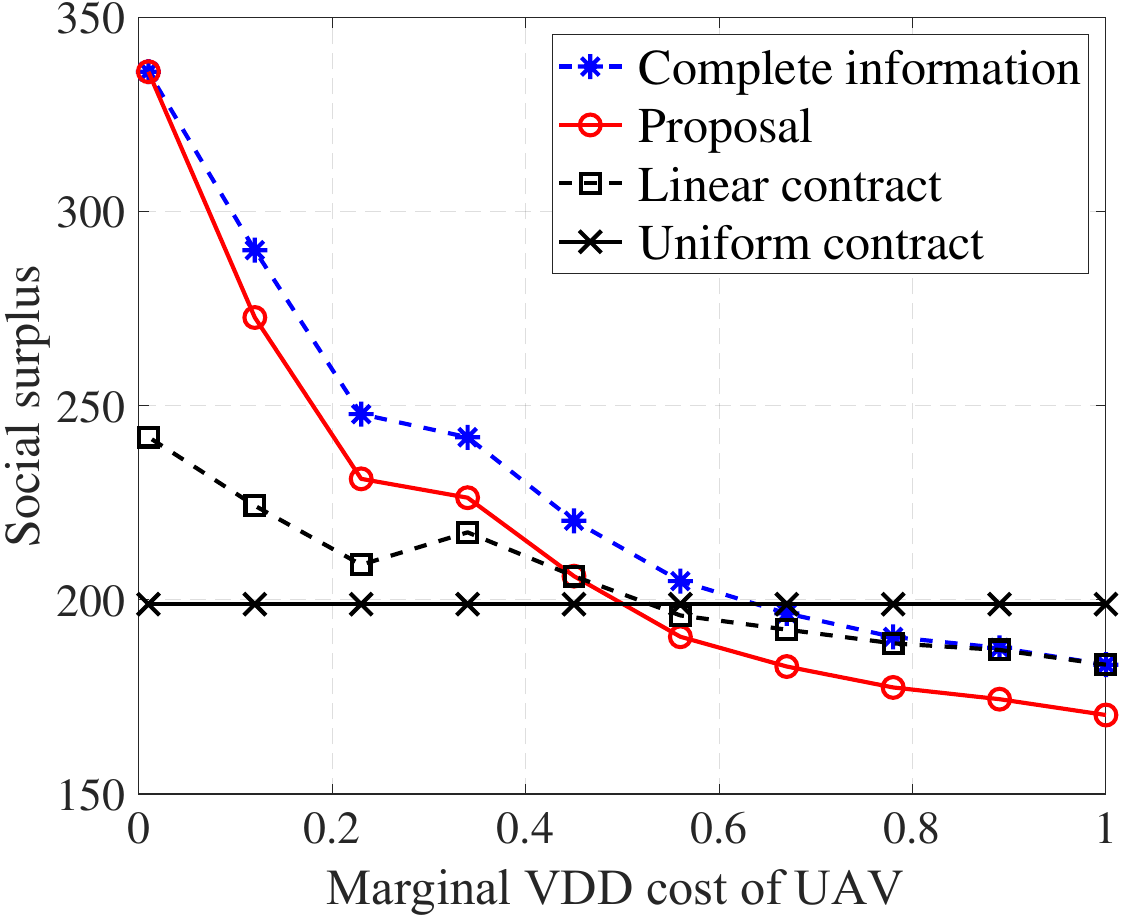}
    \caption{Social surplus vs. marginal VDD cost of UAV in the proposed scheme under partial information asymmetry, compared with other three contracts.}\label{simufig6}
\end{minipage}\vspace{-0.1cm}
\end{figure*}

\begin{figure}[t]\centering\setlength{\abovecaptionskip}{-0.02cm}
\includegraphics[height=5cm]{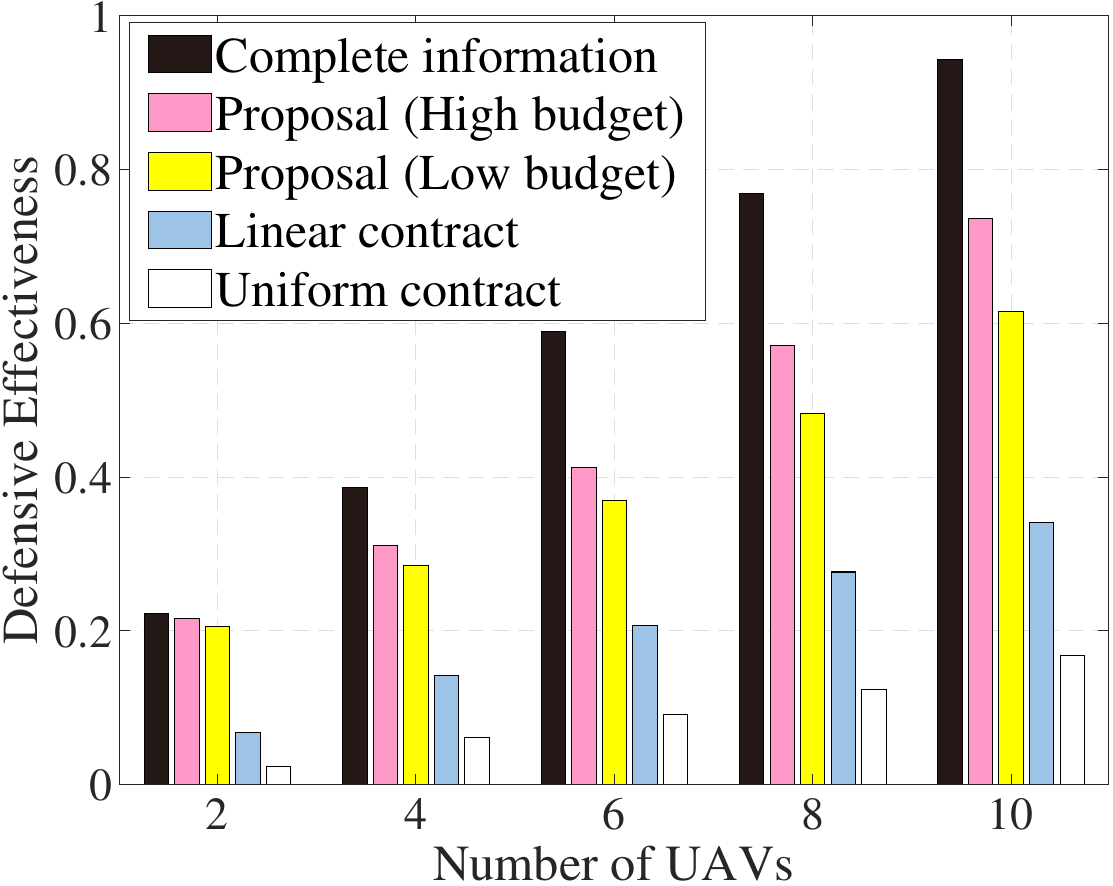}
    \caption{Defensive effectiveness vs. number of {UAVs} in the proposed scheme under partial information asymmetry, compared with other three contracts.}\label{simufig7}\vspace{-4mm}
\end{figure}

\begin{figure*}[!tbp]\setlength{\abovecaptionskip}{-0.05cm}\vspace{-0.2cm}
\begin{minipage}[t]{0.32\textwidth}
\centering
    \includegraphics[height=4.6cm,width=\textwidth]{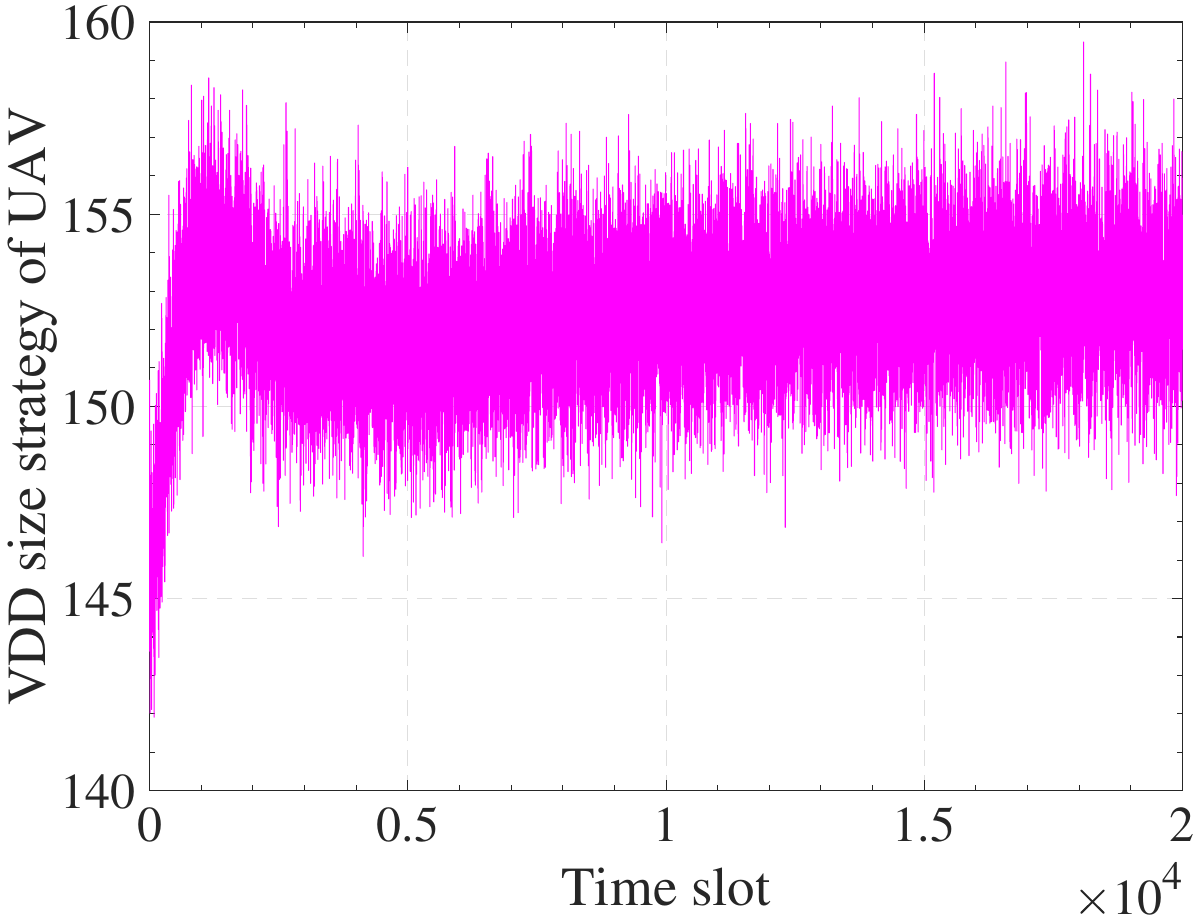}
    \caption{Evolution of UAV's strategy on VDD size using PHC learning under complete information asymmetry.}\label{simufig8}
\end{minipage}~~~~
\begin{minipage}[t]{0.32\textwidth}
\centering
    \includegraphics[height=4.6cm,width=\textwidth]{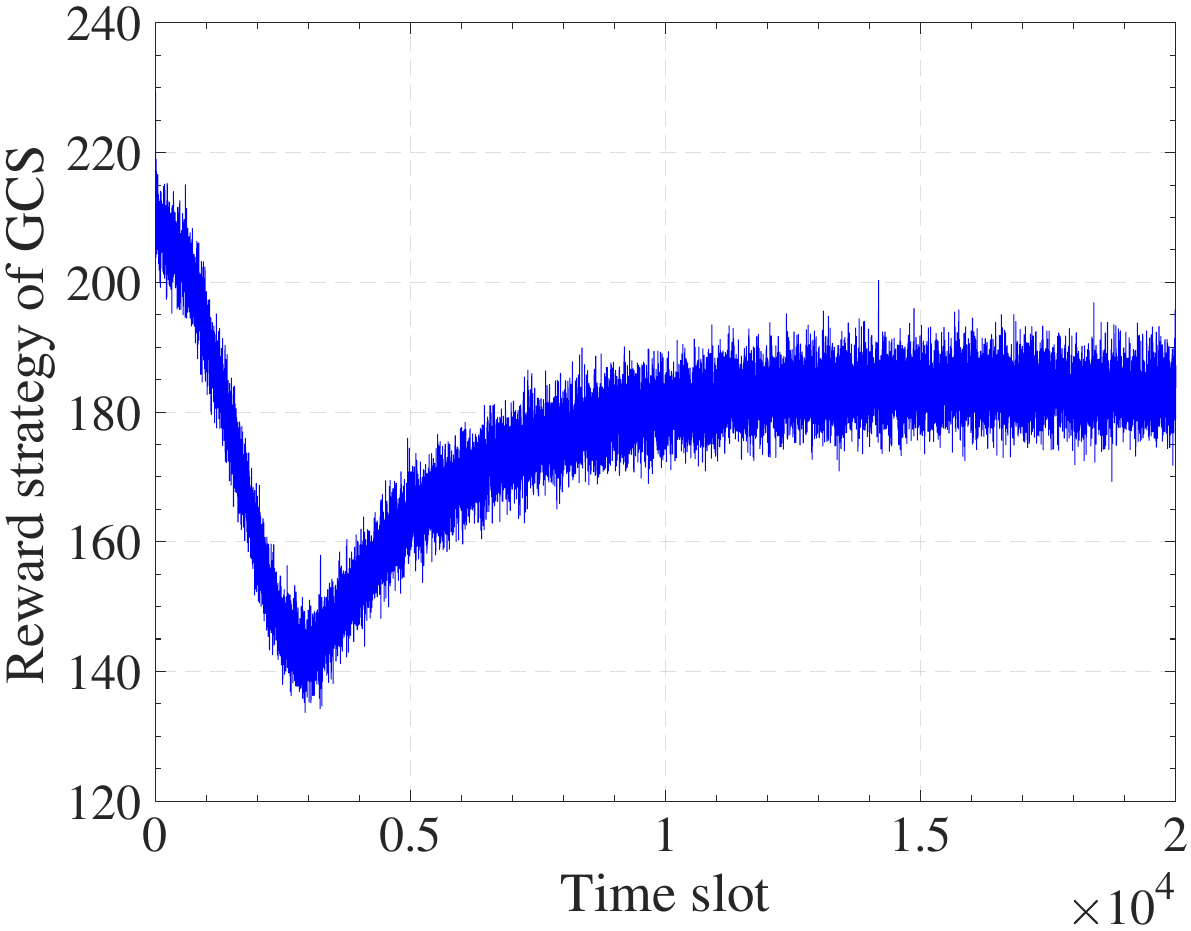}
    \caption{Evolution of GCS's reward strategy using PHC learning under complete information asymmetry.}\label{simufig9}
\end{minipage}~~~~
\begin{minipage}[t]{0.32\textwidth}
\centering
    \includegraphics[height=4.6cm,width=\textwidth]{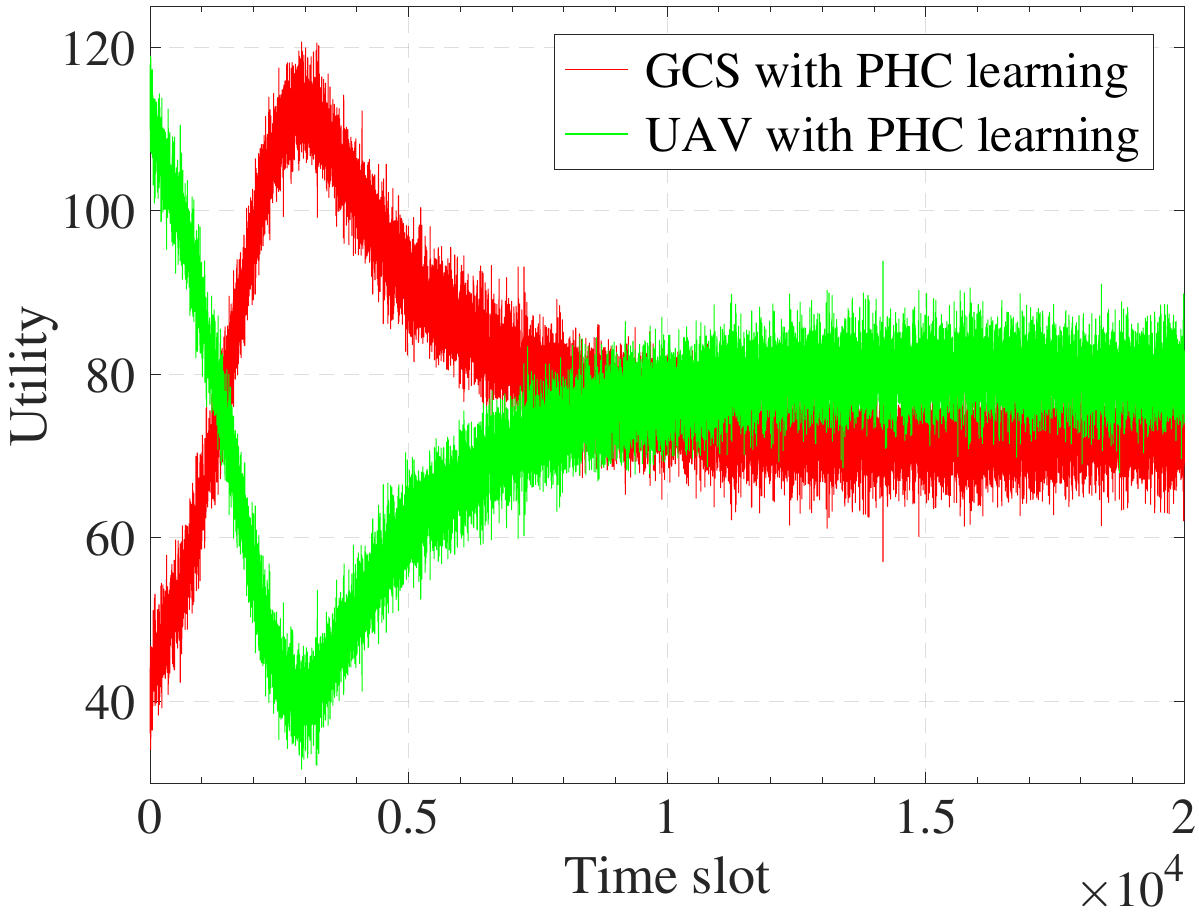}
    \caption{Evolution of utilities of UAV and GCS using PHC learning under complete information asymmetry.}\label{simufig10}
\end{minipage}\vspace{-0.2cm}
\end{figure*}

Next, we evaluate the optimal contractual VDD size and contractual reward under different schemes in Figs.~\ref{simufig1} and \ref{simufig2}, followed by the contractual feasibility analysis of the proposed scheme in Fig.~\ref{simufig3}. After that, in Figs.~\ref{simufig4}--\ref{simufig6}, we evaluate and compare the utility of the UAV, the utility of the GCS, and the social surplus in different schemes. Next, the collaborative defensive effectiveness under different schemes is evaluated in Fig.~\ref{simufig7}. Finally, we evaluate UAV's VDD size strategy, GCS's reward strategy, and their utilities during the dynamic contractual strategy-making process based on PHC in Figs.~\ref{simufig7}--\ref{simufig9}.
Here, the \emph{defensive effectiveness} is defined as $\zeta = \frac{\sum_{j\in \mathcal{J}'}{S_j}}{D_G}$,
where ${D_G}$ is the VDD requirement of the GCS. We set ${D_G}=800$ bytes.

Figs.~\ref{simufig1} and \ref{simufig2} show the optimal contractual VDD size and reward for different {types of UAVs}, respectively, in the optimal contract under partial information asymmetry. As seen in the two figures, with the increase of the marginal VDD cost of the UAV (i.e., the decrease of {UAV's type}), both the optimal contractual VDD size and reward are in decline, which accords with the monotonicity constraints in Theorem~\ref{theorem2-4}.
In addition, in the cases of information symmetry and information asymmetry, the optimal contractual VDD size is a convex function of the {UAV's} type, which is consistent with the analysis in Theorems \ref{theorem2-2} and \ref{theorem2-6}. In the linear contract, the optimal contractual VDD size and reward vary very little, given different {UAV's} types. It is because the privacy information disclosure strategy is not implemented in the linear contract, and the degree of information asymmetry cannot be reduced, resulting in the unwillingness of UAVs to contribute more local honeypot data. In the uniform contract, when the {UAV's} type changes, the optimal contractual VDD size and reward always remain the same. The reason is that the GCS only provides a single contract for all types of UAVs.

Fig. \ref{simufig3} evaluates the contractual feasibility in the proposed scheme under partial information asymmetry, by comparing the utilities of five different types of UAVs (i.e., types 1, 3, 5, 7, and 10) when selecting different contract items designed by the GCS.
It can be seen that each UAV can obtain the maximum non-negative utility only when it truthfully selects the contract designed for its type, which validates the contractual feasibility (i.e., IR and IC constraints) of the proposed optimal contract. In the proposed scheme, after each UAV truthfully chooses its contract item, the aggregated UAVs' true type information will be automatically revealed to the GCS (but the GCS still does not know that each UAV belongs to a certain type). That is to say, the optimal contract under information asymmetry enables the GCS to obtain more relevant information about UAVs' multi-dimensional private types, thereby reducing the degree of information asymmetry. In addition, in Fig.~\ref{simufig3}, when different types of UAVs select the same contract item, the higher the {UAV's} type, the greater the UAV utility. It is because when UAVs choose the same contract item, the lower the marginal UAV cost (i.e., the higher type), the higher {the} corresponding utility. Besides, as seen in Fig. \ref{simufig3}, the higher the {UAV's} type, the higher the maximum UAV utility, which conforms to Corollary~\ref{corollary1}.

Fig. \ref{simufig4} shows the UAV utility in four schemes when UAV's marginal VDD cost varies between $0.01$ and $1$.
As seen in Fig.~\ref{simufig4}, the UAV's utility remains zero under no information asymmetry. The reason is that the GCS intends to maximize its utility while enforcing IR, which is consistent with (\ref{eq:CIoptS})--(\ref{eq:CIoptR}). Moreover, in Fig.~\ref{simufig4}, the lower type brings higher utility to the UAV, which conforms to the monotonicity of the optimal contract. Overall, our proposal attains higher utility for low-type UAVs (with higher marginal VDD cost) than the linear contract, and higher utility for high-type UAVs than the uniform contract.

Fig. \ref{simufig5} shows the utility of the GCS under different marginal VDD costs of UAVs in different schemes. It can be seen that under the complete information, the GCS can obtain the highest utility among the four schemes, as the GCS fully knows the private types of all UAVs. Under the incomplete information, although the optimal contracts can motivate UAVs to select the contract items designed for their types truthfully, their true types are still unavailable to GCS. Therefore, the GCS can only approach the socially optimal utility by designing optimal contracts in the case of information asymmetry, which is consistent with Corollary~\ref{corollary0}. Similar to the above analysis in Fig. \ref{simufig4}, it can be seen from Fig.~\ref{simufig5} that in the proposed contract scheme, the higher UAV's types (i.e., with lower marginal VDD cost) can bring higher benefits (i.e., higher utility) to the GCS. Besides, we can observe that in the proposed scheme under incomplete information, the utility of the GCS is higher than that in the uniform contract, and is higher than that in the linear contract for medium and high types of UAVs. It is because the uniform contract and linear contract have no restrictions on UAV's contract selection, and cannot motivate UAVs to exhibit their true private type information, making the GCS unable to obtain higher utility.

Fig. \ref{simufig6} shows the social surplus (i.e., the sum of utilities of UAVs and the GCS) in four schemes given different UAV's marginal VDD costs. As seen in Fig. \ref{simufig6}, the utility of the UAV with the highest type (i.e., with the lowest VDD cost) in incomplete information is the same as that in the complete information, which accords with Theorems \ref{theorem2-2} and \ref{theorem2-5}. For other {UAV's} types under incomplete information, they can still obtain approximately optimal utility in the complete information. In the linear contract and uniform contract, the social surplus is generally low due to the inability to obtain additional UAV's private type information. In addition, in the uniform contract, since the GCS only provides a uniform contract item for all types of UAVs, the social surplus will not change when the {UAV's} type varies.

Fig.~\ref{simufig7} depicts the defensive effectiveness in four schemes given different number of participating UAVs. In this simulation, the budget is dynamically adjusted with the number of UAVs{,} and two types of system budgets are adopted, i.e., high budget $\Omega_1=\{160,320,480,640,800\}$ and low budget $\Omega_2=\{92,184,276,368,460\}$.
As shown in Fig.~\ref{simufig7}, our proposed scheme under partial information asymmetry outperforms both linear and uniform contracts in terms of higher defensive effectiveness, and its gap with the ideal complete information contract shrinks as the number of UAVs decreases. The reason is that the reward in the uniform contract and linear contract is either fixed or linear with the VDD. Notably, the relationship between the optimal reward and optimal VDD size in optimal contracts is nonlinear in our proposal, creating a stronger incentive for UAVs to contribute more VDD and improve defensive effectiveness. Besides, in our proposal, the higher system budget results in better defensive performance. It can be explained as follows. According to Lemma 2, the GCS tends to exhaust the available budget. Moreover, according to Theorem 4, a higher budget can incentivize UAVs' high amount of contributed VDD, thereby leading to improved defensive effectiveness.

Next, in Figs.~\ref{simufig8}--\ref{simufig10}, we show the convergence of PHC-based optimal dynamic contract for a randomly selected UAV under complete information asymmetry. The evolutions of UAV's VDD size strategy and GCS's reward strategy via PHC learning are shown in Fig.~\ref{simufig8} and Fig.~\ref{simufig9}, respectively. The evolutions of average utilities of the UAV and the GCS in PHC learning are shown in Fig. \ref{simufig10}.
As seen in these three figures, both the VDD size and reward in dynamic contracts can converge to stable and optimal values, validating the feasibility of the proposed two-layer PHC learning-based incentive mechanism. In Fig.~\ref{simufig8}, the VDD size first increases then converges to a stable state, while the corresponding contractual reward in Fig.~\ref{simufig9} first decreases then grows to reach the stable state. In Fig.~\ref{simufig10}, the utility of UAV first decreases then grows to reach the stable value, while the utility of GCS first increases then gradually drops to the convergent value.
The reasons are as follows. Motivated by the initial high reward of the GCS, the UAV intends to share more VDD to improve its utility. Then, after observing UAV's high VDD contribution, the GCS gradually decreases its reward to increase its utility. After that, the UAV and GCS continuously pursue their maximized utilities by seeking the optimal VDD size and reward strategy based on their observed system states.

\section{Conclusion}\label{sec:CONSLUSION}
In this paper, we have proposed an optimal and feasible incentive mechanism to promote collaborative defense for UAVs by sharing their captured VDD in local honeypots.
Firstly, a novel honeypot game has been formulated between the GCS and UAVs with distinct types (i.e., VDD cost and communication delay), the solution of which is to design optimal VDD-reward contracts under both partial and complete information asymmetry scenarios.
Secondly, we have analytically derived the optimal contracts with budget and contract feasibility under partial information asymmetry, by summarizing UAV's multi-dimensional private information into a one-dimensional metric.
Thirdly, a two-layer PHC learning algorithm has been devised to intelligently address the dynamic contract design problem under complete information asymmetry and time-varying UAV networks. Numerical results have demonstrated that the proposed scheme can effectively encourage UAVs to share local VDD with the GCS and effectively enhance UAV's utility and collaborative defensive performance under both partial and complete information asymmetry.
For future work, we plan to investigate the federated learning approaches for privacy-preserving honeypot data sharing and defense service offerings among UAVs. Besides, the trust-free honeypot data sharing based on lightweight blockchain will be further studied.

\begin{appendices}
\section{Proof of Lemma 1}\label{Appendix A}
\textbf{Lemma 1}.
For any VDD data size $S_{j}\in[0, S_{\max}]$, the optimal reward strategy {of} the GCS is:
\begin{align}\label{eq:Complete-OptimalPrice}
     R_{j}^{*}\left(S_{j}\right) =\left\{ \begin{array}{ll}
     0,&\forall j \notin \mathcal{J}';\\
     {C_j} S_{j} + C_0,&\forall j \in \mathcal{J}'.\\
     \end{array} \right.
     \end{align}

\begin{IEEEproof}
Obviously, the optimal payment reward {given by} the GCS is zero for non-participating UAVs. For the optimal rewards of participating UAVs, we prove it by contradiction. Suppose that there exists an optimal reward policy $\hat{R}_{j}$ that satisfies $\hat{R}_{j}-{C_j} S_{j} - C_0 \neq 0$. First, we assume that the optimal reward policy satisfies $\hat{R}_{j}-{C_j} S_{j} - C_0 < 0$, which contradicts the IR constraint. Second, we suppose it satisfies $\hat{R}_{j}-{C_j} S_{j} - C_0 > 0$. Since the utility of the GCS decreases as the payment reward increases, the GCS can continuously increase its utility by reducing the reward $\hat{R}_{j}$ until $\hat{R}_{j }-{C_j} S_{j} - C_0 = 0$, which contradicts the assumption that $\hat{R}_{j}-{C_j} S_{j} - C_0 > 0$. To sum up, there exists no optimal reward strategy $\hat{R}_{j}$ that satisfies $\hat{R}_{j}-{C_j} S_{j} - C_0 \neq 0$.
\end{IEEEproof}

\section{Proof of Lemma 2}\label{Appendix B}
\textbf{Lemma 2}.
The BF constraint in (12) can be simplified as:
\begin{align}\label{eq:BF-simp}
\sum\nolimits_{j=1}^{J'} {N_j R_j} = \Omega.
\end{align}

\begin{IEEEproof}
Suppose that $\sum\nolimits_{j=1}^{J'} {N_j R_j} < \Omega$. Then, the GCS could always prefer a larger $R_j$, which allows for a larger VDD size $S_j$ from type-$j$ UAV, to enhance the defensive performance until $\sum\nolimits_{j=1}^{J'} {N_j R_j} = \Omega$.
\end{IEEEproof}

\section{Proof of Theorem 1}\label{Appendix C}
\textbf{Theorem 1}.
Under the complete information scenario, the {contractual VDD size and contractual reward for each type of UAV} in the optimal contract {$\Phi^*=\{T_{\max},\{S_j^*,R_j^*\}_{j\in \mathcal{J}}\}$} are:
\begin{enumerate}
   \item $\forall j \notin \mathcal{J}'$, $S_{j}=R_{j}=0$.
   \item $\forall j \in \mathcal{J}'$, we have
     \begin{numcases}{}	
     S_{j}^{*}\!=\!\min\left\{S_{\max},\max\left\{\frac{\Lambda}{T_j C_j} -1 ,0\right\}\right\}, \label{eq:CIoptS} \hfill \\
     R_{j}^{*}= {C_j} S_{j}^{*} + C_0, \label{eq:CIoptR} \hfill
 \end{numcases}
where {$\Lambda$ is the abbreviation for}
\begin{align}\label{eq:Lambda}
\Lambda = \frac{\Omega + \sum\nolimits_{j=1}^{J'}{N_j {C_j}}- C_0\sum\nolimits_{j=1}^{J'}{N_j}}{\sum\nolimits_{j =1}^{J'}{\frac{N_j}{T_j}}}.
\end{align}
\end{enumerate}

\begin{IEEEproof}
Obviously, for any non-participating UAV {with type} $j \notin \mathcal{J}'$, the optimal {contractual} VDD size and payment are equal to zero. {The optimal contract for the non-participating UAVs is $\Phi _{j}^{*}=\{S_{j}^{*},R_{j}^{*}\}=\left\{ 0,0 \right\} ,\ \forall j\ni \mathcal{J}'$, and the corresponding GCS's utility $\mathcal{U}_{G}(\Phi _{j}^{*})$ equals to zero. As such, we only need to consider the optimal contract problem for participating UAVs with type $j \in \mathcal{J}'$. In other words, we only need to consider the case that $1_{T_j\le T_{\max}}=1$. For participating UAVs, the objective function can be rewritten as $\mathcal{U}_G =\sum_{j\in \mathcal{J}'}{\varpi}\frac{N_j}{T_j}\log \left( 1+S_j \right) -N_jR_j$.}

For any participating UAV {with type} $j \in \mathcal{J}'$, {according to the Lagrange function ${\mathscr{L}}(S_j,\lambda_1)$ defined in (20) in the main text, we have} $\frac{\partial^2{\mathscr{L}}(S_j,\lambda_1)}{\partial S_{j}^2}=-\frac{\varpi N_j}{{T_j}\left( 1+ S_{j} \right) ^2}<0$. It indicates that ${\mathscr{L}}(S_j,\lambda_1)$ is {strictly convex} about $S_{j}$. Therefore, {according to the differential and integral calculus,} the optimal contractual VDD size $S_{j}^*$ can be found {(i) at the point where} $\frac{\partial {\mathscr{L}}(S_j,\lambda_1)}{{\partial S_{j}}}= 0$ and $\frac{\partial {\mathscr{L}}(S_j,\lambda_1)}{{\partial \lambda_1 }}= 0$ {meet simultaneously or (ii)} at the boundary point. Hence, $S_{j}^{*}=\min\big\{S_{\max},\max\big\{\tilde{S}_j^*,0\big\} \big\}$.

Let $\frac{\partial {\mathscr{L}}(S_j,\lambda_1)}{{\partial S_{j}}}= 0$ and $\frac{\partial {\mathscr{L}}(S_j,\lambda_1)}{{\partial \lambda_1 }}= 0$, {we have
\begin{align}
&\frac{\partial {\mathscr{L}}(S_j,\lambda_1)}{{\partial S_{j}}} = 0 \Rightarrow \frac{\varpi}{{T_j}(1 \!+\! S_{j})} + (\lambda_1 - 1)C_j = 0 \label{eq:deriapp1} \\
&\frac{\partial {\mathscr{L}}(S_j,\lambda_1)}{{\partial \lambda_1 }}= 0 \Rightarrow \sum\nolimits_{j=1}^{J'} {N_j \left({C_j} S_{j} + C_0 \right)} = \Omega. \label{eq:deriapp2}
\end{align}
From (\ref{eq:deriapp1}), we can obtain $\tilde{S}_j^* =\frac{\varpi}{1- \lambda_1}\cdot \frac{1}{T_j C_j} -1$. By substituting $\tilde{S}_j^*$ with ${S}_j$ in (\ref{eq:deriapp2}), after some derivations and simple transformations, we can obtain
\begin{align}
\frac{\varpi}{1- \lambda_1} = \frac{\Omega + \sum\nolimits_{j=1}^{J'}{N_j {C_j}}- C_0\sum\nolimits_{j=1}^{J'}{N_j}}{\sum\nolimits_{j =1}^{J'}{\frac{N_j}{T_j}}} \triangleq \Lambda.
\end{align}
%After some derivations and simple transformations, we can obtain
As such,} the optimal VDD size strategy is
\begin{align}
\tilde{S}_j^*&=\frac{1}{T_j C_j}\cdot \frac{\Omega + \sum\nolimits_{j=1}^{J'}{N_j {C_j}}- C_0\sum\nolimits_{j=1}^{J'}{N_j}}{\sum\nolimits_{j =1}^{J'}{\frac{N_j}{T_j}}} - 1 \nonumber \\
&=\frac{\Lambda}{T_j C_j} -1.
\end{align}
According to (\ref{eq:Complete-OptimalPrice}) in Lemma 1, the corresponding optimal reward strategy can be derived, as shown in (\ref{eq:CIoptR}).
\end{IEEEproof}

\section{Proof of Theorem 2}\label{Appendix A}
\textbf{Theorem 2}.
A contract $\Phi=\{T_{\max},\{\Phi_j\}_{j\in \mathcal{J}}\}$ is feasible if and only if it meets the following conditions:
\begin{enumerate}
  \item $\forall j \notin \mathcal{J}'$, $S_{j}=R_{j}=0$.
  \item $\forall j \in \mathcal{J}'$, the following three conditions hold:
  \begin{numcases}{}	
     0\le S_1 \le \cdots \le S_{J'} \& 0\le R_1 \le \cdots \le R_{J'},~~~~\ \label{eq:theo1cons1} \hfill \\
     R_1-{C_1}S_1-{C_0}\ge 0,  \label{eq:theo1cons2} \hfill \\
     \begin{gathered}
     C_{j}\left( S_j - S_{j-1} \right) \le R_{j} - R_{j-1}~~~~~~~~~~~~~~~~~~~ \\~~~~~~~~\le C_{j-1}\left( S_{j}-S_{j-1} \right),\ j = 2,\cdots,{J'}. \label{eq:theo1cons3} \hfill
     \end{gathered}
 \end{numcases}
\end{enumerate}

\begin{IEEEproof}
Obviously, in case $1$ (i.e., $j \notin \mathcal{J}'$), the corresponding contractual VDD size and reward are zero (i.e., $S_{j}=R_{j}=0$). Thereinafter, we focus on the case $2$ (i.e., $j \in \mathcal{J}'$) for UAV types that can timely transmit their VDD to the GCS.
As the contractual feasibility means that both IR and IC constraints are satisfied, we need to prove the equivalence between the constraints (\ref{eq:theo1cons1})--(\ref{eq:theo1cons3}) and the IR\&IC constraints in (13)--(14).

\emph{1) Necessity: }
We need to prove that if IR and IC constraints hold for all types of UAVs, then the constraints (\ref{eq:theo1cons1})--(\ref{eq:theo1cons3}) automatically hold.

(i) According to IR constraint for type-$1$ UAV, we have $R_1-{C_1} S_1 -C_0\ge 0$, which is shown in (\ref{eq:theo1cons2}).

(ii) According to IC constraints for type-$j$ and type-$k$ UAVs ($j\ne k$), we have
\begin{align}\label{eq:ii-1}
R_{j}-C_j S_{j}\ge R_{k}-C_j S_{k},
\end{align}
\begin{align}\label{eq:ii-2}
R_{k}-C_k S_{k}\ge R_{j}-C_k S_{j}.
\end{align}
Combining the above two constraints, we can derive $(C_j-C_k)(S_j-S_k)\le 0$. As $C_1>C_2>\cdots>C_{J'}$ and $S_j \ge 0$, we have $0\le S_1 \le S_2 \le\cdots \le S_{J'}$. Besides, based on (\ref{eq:ii-1}), we have
\begin{align}\label{eq:}
C_j(S_j-S_k) \le R_j-R_k \le C_k(S_j-S_k),
\end{align}
which leads to $0\le R_1 \le R_2 \le \cdots \le R_{J'}$. Hence, we obtain the monotonicity constraint (\ref{eq:theo1cons1}).

(iii) Considering IC constraints for two neighboring contract items, we have
\begin{align}\label{eq:iii-1}
R_{j}-C_j S_{j}\ge R_{j-1}-C_j S_{j-1},
\end{align}
\begin{align}\label{eq:iii-2}
R_{j-1}-C_{j-1} S_{j-1}\ge R_{j}-C_{j-1} S_{j}.
\end{align}
Combining the above two constraints, we can obtain $C_{j}\left( S_{j}-S_{j-1} \right) \le R_{j}-R_{j-1}\le C_{j-1}\left( S_{j}-S_{j-1} \right)$, which is shown in (\ref{eq:theo1cons3}).

To summarize, (\ref{eq:theo1cons1})--(\ref{eq:theo1cons3}) are the necessity conditions of IR and IC constraints.

\emph{2) Sufficiency: }We prove by mathematical induction that if the constraints (\ref{eq:theo1cons1})--(\ref{eq:theo1cons3}) hold, then both IR and IC constraints are met for all UAV types.
Let $\mathcal{A}(q)$ denote a subset of $\Phi$, which consists of the first $q$ contract items in $\Phi$, i.e., $\mathcal{A}(q) = \{(S_{j},R_{j})|j=1,2,\cdots,q\}$. Let $\mathcal{J}(q)=\{1,2,\cdots,q\}$.
When $q=1$, as there exists only one UAV type, only the IR constraint needs to be considered for a feasible contract. Obviously, according to (\ref{eq:theo1cons2}), we have $R_1-C_1 S_1 - C_0\ge 0$. Then, $\mathcal{A}(1)$ is proved to be feasible.

Next, we show that if $\mathcal{A}(q)$ is feasible, $\mathcal{A}(q+1)$ is also feasible. To achieve this goal, we need to prove the following two aspects.
(i) Both the IR and IC constraints are met for the new type $q+1$, i.e.,
\begin{numcases}{}
  R_{q+1}-C_{q+1} S_{q+1}\ge 0, \label{eq:si-01}\hfill \\
  R_{q+1}-C_{q+1} S_{q+1}\ge R_{j}-C_{q+1} S_{j},\forall j \!\in\! \mathcal{J}(q). \label{eq:si-02}\hfill
\end{numcases}
And (ii) for all existing UAV types $j\in \mathcal{J}(q)$, IC constraints are met in the existence of type $q+1$, i.e.,
\begin{align}\label{eq:a1-6}
R_{j}-C_{j} S_{j}\ge R_{q+1}-C_{j} S_{q+1},\forall j \in \mathcal{J}(q).
\end{align}

\emph{Proof of part (i):} Due to the feasibility of $\mathcal{A}(q)$, the IC constraint for type-$q$ UAV is satisfied for any $k \in \mathcal{A}(q)$:
\begin{align}\label{eq:a1-7}
R_q-C_q S_q\ge R_{k}-C_q S_{k}.
\end{align}
According to the left part of constraint (\ref{eq:theo1cons3}), we can attain
\begin{align}\label{eq:a1-8}
R_{q+1}\ge R_q + C_{q+1}(S_{q+1} - S_q).
\end{align}
Combining the above two inequalities, we can obtain
\begin{align}\label{eq:a1-9}
R_{q+1}-C_{q+1} S_{q+1} &\ge R_{q}-C_{q+1} S_{q} \nonumber \\
&\ge R_{k} - C_{q+1} S_{q} + C_{q} (S_{q}-S_{k}) \nonumber \\
&\ge R_{k} - C_{q+1} S_{q} + C_{q+1} (S_{q}-S_{k}) \nonumber \\
&= R_{k} - C_{q+1} S_{k},\forall k \in \mathcal{A}(q).
\end{align}
Thereby, the IC constraint is satisfied for type-$(q+1)$ UAV.

As IR constraints hold for all type-$k$ UAVs, we can further obtain $R_{k} - C_{k} S_{k} - C_0 \ge 0$, $\forall k \in \mathcal{A}(q)$.
Beside, since $k< q+1$, we have $C_k> C_{q+1}$. As such, we have
\begin{align}\label{eq:a1-8-2}
R_{q+1}-C_{q+1} S_{q+1} &\ge R_{k} - C_{q+1} S_{k} \nonumber \\
&\ge R_{k} - C_{k} S_{k}, \forall k \in \mathcal{A}(q) \nonumber \\
&\ge 0.
\end{align}
According to (\ref{eq:a1-8-2}), the IR constraint is satisfied for type-$(q+1)$ UAV. Hence, part (i) is proved.

\emph{Proof of part (ii):} Since $\mathcal{A}(q)$ is feasible, the IC constraint holds $\forall k \in \mathcal{A}(q)$:
\begin{align}\label{eq:a1-10}
R_k-C_k S_k\ge R_{q}-C_k S_{q}.
\end{align}
According to the right part of constraint (\ref{eq:theo1cons3}), we can attain
\begin{align}\label{eq:a1-11}
R_{q+1}\le R_{q} + C_{q}(S_{q+1} - S_{q}).
\end{align}
Combining the above two inequalities, we can obtain
\begin{align}\label{eq:a1-12}
R_{k}-C_{k} S_{k} &\ge R_{q+1} - C_{k} S_{q} - C_{q}(S_{q+1} - S_{q})\nonumber \\
&\ge R_{q+1} - C_{k} S_{q} - C_{k}(S_{q+1} - S_{q}) \nonumber \\
&=  R_{q+1} - C_{k}S_{q+1},\forall k \in \mathcal{A}(q).
\end{align}
Hence, part (ii) is proved.
In summary, we have proved that (a) $\mathcal{A}(1)$ is feasible, and (b) if $\mathcal{A}(q)$ is feasible, $\mathcal{A}(q+1)$ is also feasible. Based on the mathematical induction method, it can be concluded that $\mathcal{A}= \mathcal{A}(J')$ is feasible. Theorem $2$ is proved.
\end{IEEEproof}

\section{proof of Theorem 3}\label{Appendix B}
\textbf{Theorem 3}.
Given any VDD volume sequence $\mathbf{S} = \{S_{j}\}_{j \in \mathcal{J}'}$ meeting $0\le S_{1} \le \cdots \le S_{J'} \le S_{\max}$, the unique optimal reward strategy $\mathbf{R}^*=\{R_{j}^*\}_{j \in \mathcal{J}'}$ is attained by:
\begin{enumerate}
  \item $\forall j \notin \mathcal{J}'$, $R_{j}^*(\mathbf{S})=0$.
  \item $\forall j \in \mathcal{J}'$, we have
  \begin{align}\label{eq:optreward1}
    R_{j}^{*}\left( \mathbf{S} \right) =\left\{ \begin{array}{l}
    	R_{j-1}^{*}\left( \mathbf{S} \right) +C_j\left( S_{j}-S_{j-1} \right) ,\\ ~~~~~~~~~~~~~~~~~~ j=2,...,J';\\
    	{C_j} S_{j} + C_0,\ ~~~~j=1.\\
    \end{array} \right.
  \end{align}
\end{enumerate}

\begin{IEEEproof}
Obviously, in case $1$, for UAVs in $\mathcal{J} \backslash \mathcal{J}'$, the optimal contractual reward equals to zero. In what follows, we prove the case $2$ by contradiction for UAVs in $\mathcal{J}'$.

\emph{1) Optimality:} Notably, the reward strategy in (\ref{eq:optreward1}) meets the constraints (12) and (13) in Theorem~1, and it satisfies the monotonicity constraint in (11) under the monotonic VDD size strategy. Here, we prove that the reward strategy in (\ref{eq:optreward1}) maximizes the GCS's utility. Given the fixed VDD size strategy $\mathcal{S}$, the maximum utility of the GCS in (6) can be acquired by minimizing the $\sum\nolimits_{j=1}^{J'}{{N_j}{R_{j}}}$.
Suppose that there exists a reward sequence $\widehat{\mathbf{R}} = \{\hat{R}_{j}\}_{j \in \mathcal{J}'}$ such that $\sum\nolimits_{j=1}^{J'}{{N_j}{\hat{R}_{j}}}< \sum\nolimits_{j=1}^{J'}{{N_j}{{R}_{j}^*}}$. As a consequence, there exists at least one reward $\hat{R}_{j} < R_{j}^{*}$.
According to Theorem~1, to ensure the contractual feasibility, $\widehat{\mathbf{R}}$ should satisfy:
\begin{align}\label{eq:a2-1}
\hat{R}_{j-1} + C_j (S_j - S_{j-1}) \le \hat{R}_{j} < R_{j}^{*}.
\end{align}
According to (\ref{eq:optreward1}), the above inequality in (\ref{eq:a2-1}) can be reformulated as:
\begin{align}\label{eq:a2-2}
\hat{R}_{j-1} <  R_{j}^{*} - C_j (S_j - S_{j-1}) = R_{j-1}^{*}.
\end{align}
Continuing the above process until $j=1$, we can eventually obtain that $\hat{R}_{1} < {R}_{1}^* = {C}_{1} {S}_{1} + C_0$, which violates the IR constraint for type-$1$ UAVs. Thereby, there does not exist any feasible reward strategy $\widehat{\mathbf{R}}$, and the utility of the GCS is optimized by applying the reward strategy in (\ref{eq:optreward1}).

\emph{2) Uniqueness:} To prove the uniqueness of the optimal reward strategy in (\ref{eq:optreward1}), we first assume that there exists a reward strategy $\widehat{\mathbf{R}} = \{\hat{R}_{j}\}_{j \in \mathcal{J}'} \ne \mathbf{R}^*$ such that $\sum\nolimits_{j=1}^{J'}{{N_j}{\hat{R}_{j}}}= \sum\nolimits_{j=1}^{J'}{{N_j}{R_{j}^*}}$. Hence, there must exist at least one reward $\hat{R}_{j} \ne R_{j}^{*}$. Without loss of generality, it is assumed that $\hat{R}_{j} > R_{j}^{*}$. As such, there must exist another reward $\hat{R}_{k} < R_{k}^{*}$. Using the same method, we obtain a contradiction, which implies that the optimal reward strategy in (\ref{eq:optreward1}) is unique. Theorem $3$ is proved.
\end{IEEEproof}

\section{proof of Theorem 4}\label{Appendix C}
\textbf{Theorem 4}.
Under partial information asymmetry, the optimal contractual VDD size strategy to solve the {relaxed} Problem~2-1 without constraint C1 is attained as:
\begin{align}\label{eq:OptimalSize1}
S_{j}^*=\min\left\{S_{\max},\max\left\{\frac{N_j }{A_j T_j}\cdot \mathfrak{R} -1 ,0\right\}\right\},
\end{align}
where $\mathfrak{R}$, $A_j$, and $\Delta C_j$ are defined as follows:
\begin{align}\label{eq:R2}
\mathfrak{R} = \frac{\Omega + \sum\nolimits_{j=1}^{J'}{A_j}- C_0\sum\nolimits_{j=1}^{J'}{N_j}}{\sum\nolimits_{j =1}^{J'}{\frac{N_j}{T_j}}}.
\end{align}
\begin{align}\label{eq:Aj}
A_{j}\!=\!\left\{ \begin{array}{l}
 {N_{J'} C_{J'}},~~~~~~~~~~~~~~~~~~~j\!=\!J';\\
 {N_j} C_j + \Delta C_j \sum\limits_{k=j+1}^{J'}{N_k},~ j\!\le\! J'-1.\\ %j\!=\!1,\!\cdots\!,J'\!-\!1
\end{array} \right.
\end{align}
\begin{align}\label{eq:deltaCj}
\Delta C_j = C_j -C_{j+1}.
\end{align}

\begin{IEEEproof}
By substituting the optimal reward strategy $R_j^*(\mathbf{S})$ in (33) into the BF constraint C4 in (32), we have
\begin{align}\label{eq:a3-2}
&\Omega =\sum\limits_{j \in \mathcal{J}'} {N_j R_j^*} = N_1 R_1^* + \sum\limits_{j =2}^{J'} {N_j R_j^*} \nonumber \\
&= \sum\limits_{j=1}^{J'}{N_j {C_j} S_{j}} +  \sum\limits_{j=2}^{J'}{N_j{\sum\limits_{k=1}^{j-1}{(C_k-C_{k+1})S_{k}}}} + \sum\limits_{j=1}^{J'}{N_j {C_0}} \nonumber \\
&= \sum\limits_{j=1}^{J'}{N_j {C_j} S_{j}} +  \sum\limits_{j=1}^{J'-1}{{(C_j-C_{j+1})S_{j}}{\sum\limits_{k=j+1}^{J'}{N_k}}} + \sum\limits_{j=1}^{J'}{N_j {C_0}} \nonumber \\
&= \sum\limits_{j=1}^{J'}{N_j {C_j} S_{j}} +  \sum\limits_{j=1}^{J'-1}{{\Delta C_j S_{j}}{\sum\limits_{k=j+1}^{J'}{N_k}}} + \sum\limits_{j=1}^{J'}{N_j {C_0}}\nonumber \\
&= \sum\limits_{j=1}^{J'}\Big({{A_j} S_{j} + N_j {C_0}}\Big).
\end{align}
Besides, the GCS's utility function $\mathcal{U}_G(\Phi)$ in (6) can be rewritten as
\begin{align}\label{eq:a3-1}
\mathcal{U}_G(\Phi)=\mathcal{U}_{G}(S_j)= \sum\limits_{j \in \mathcal{J}'} {\varpi \frac{N_j}{T_j}\log \left(1 +  S_j \right)} - \Omega.
\end{align}

Hence, for the relaxed Problem 2-1 without the monotonicity constraint C1, the corresponding Lagrangian function can be expressed as:
\begin{align}\label{eq:Lagrangian2}
{\mathscr{L}}(S_j,\lambda_2) &= \mathcal{U}_{G}(S_j) + \lambda_2 \Big(\sum\limits_{j=1}^{J'}\Big({{A_j} S_{j} + N_j {C_0}}\Big) - \Omega\Big) \nonumber \\
&=\sum\limits_{j =1}^{J'} {\frac{\varpi N_j}{T_j}\log \left(1 \!+\! S_j \right)} + \lambda_2 \sum\limits_{j=1}^{J'}\Big({{A_j} S_{j} + N_j {C_0}}\Big) \nonumber \\
&~~~- (\lambda_2 + 1) \Omega,
\end{align}
where $\lambda_2$ represents the Lagrange multiplier.

As $\frac{\partial^2{\mathscr{L}}(S_j,\lambda_2)}{\partial S_{j}^2}=-\frac{\varpi N_j}{{T_j}\left( 1+S_{j} \right) ^2}<0$, ${\mathscr{L}}(S_j,\lambda_2)$ is strictly concave with respect to $S_{j}$. Thereby, the optimal VDD size strategy $S_{j}^*$ can be obtained by
\begin{align}\label{eq:OptimalSize1}
S_{j}^*=\min\left\{S_{\max},\max\left\{\tilde{S}_{j}^{*},0\right\}\right\},
\end{align}
where the point $\tilde{S}_{j}^{*}$ simultaneously satisfies $\frac{\partial {\mathscr{L}}(S_j,\lambda_2)}{{\partial S_{j}}}= 0$ and $\frac{\partial {\mathscr{L}}(S_j,\lambda_2)}{{\partial \lambda_2 }}= 0$.
After some derivations and simple transformations, we can obtain
\begin{align}
\tilde{S}_j^*&=\frac{N_j}{A_j T_j}\cdot \frac{\Omega + \sum\nolimits_{j=1}^{J'}{A_j}- C_0\sum\nolimits_{j=1}^{J'}{N_j}}{\sum\nolimits_{j =1}^{J'}{\frac{N_j}{T_j}}} -1\nonumber \\
&=\frac{N_j}{A_j T_j}\cdot \mathfrak{R} -1.
\end{align}
Theorem 4 is proved.
\end{IEEEproof}

\end{appendices}

%\vspace{-0.09cm}
\bibliographystyle{IEEETran}
\bibliography{ref-UAVhoneypotTrans}

\end{document}